\documentclass[
  superscriptaddress,
  aps,
  prd,
  twocolumn,                 
  preprintnumbers,
  amsmath,
  amssymb,
  nofootinbib,
  longbibliography
]{revtex4-1}
\pdfoutput=1

\usepackage[utf8]{inputenc}
\usepackage[T1]{fontenc}
\usepackage{microtype}        

\usepackage{graphicx}
\usepackage{mathtools}        
\usepackage{amsthm}
\usepackage{bm}               
\usepackage{bbm}              
\usepackage{array}
\usepackage{tabularx}
\usepackage{booktabs}
\usepackage{multirow}
\usepackage{graphicx}
\usepackage{wrapfig}          

\usepackage{xcolor}
\definecolor{nicered}{rgb}{0.7,0.1,0.1}
\definecolor{nicegreen}{rgb}{0.1,0.5,0.1}
\usepackage[colorlinks=true,linkcolor=nicered,citecolor=blue,urlcolor=blue]{hyperref}

\usepackage{enumitem}
\usepackage{verbatim}
\usepackage{tikz}
\usepackage{stackengine}
\usepackage{cancel}
\usepackage[normalem]{ulem}    

\usepackage{setspace}

\numberwithin{equation}{section}

\newcommand{\be}{\begin{equation}}
\newcommand{\ee}{\end{equation}}
\newcommand{\bea}{\begin{aligned}} 
\newcommand{\eea}{\end{aligned}}

\allowdisplaybreaks

\begin{document}
\linespread{1.28}\selectfont

\title{Reviving sub-keV warm dark matter: a UVLF-based analysis}

\author{Raymond T. Co}
\thanks{rco@iu.edu}
\affiliation{Physics Department, Indiana University, Bloomington, IN 47405, USA}
\author{Siu Cheung Lam}
\thanks{siuclam@iu.edu}
\affiliation{Physics Department, Indiana University, Bloomington, IN 47405, USA}
\author{Sai Chaitanya Tadepalli}
\thanks{\textbf{saictade@iu.edu} (for correspondence) }
\affiliation{Physics Department, Indiana University, Bloomington, IN 47405, USA}
\author{Tomo Takahashi}
\thanks{tomot@cc.saga-u.ac.jp}
\affiliation{Department of Physics, Saga University, Saga 840-8502, Japan \vspace{1cm}}

\begin{abstract}
Thermal warm dark matter (WDM) particles with $m_{\rm WDM} \leq 1~\mathrm{keV}$ are ruled out at more than $4\sigma$ by multiple observational probes, owing to the strong suppression of small-scale structure induced by early-time free-streaming. Recently, it was highlighted that a small admixture of
$\sim1\%$ ($f_{\rm CDM} \sim\!0.01$) cold dark matter (CDM) endowed with a blue-tilted isocurvature power spectrum could offset the WDM-induced suppression and relax the WDM mass bound by a factor of $\mathcal{O}(10)$. If viable, this ``warm + cold-isocurvature'' scenario would allow sub-keV WDM particles to constitute nearly the full dark matter abundance while potentially alleviating some small-scale tensions. In this work, we test this mechanism by constraining the WDM mass $m_{\rm WDM}$ while marginalizing over CDM isocurvature parameters. We combine ultraviolet luminosity function measurements from the \textit{Hubble Space Telescope} and \textit{James Webb Space Telescope} over redshift $4 \leq z \leq 11$ with CMB, BAO, and SNe data. For a pure WDM model, our joint analysis yields a lower bound $m_{\rm WDM} > 1.8~\mathrm{keV}$ (95\%~credible intervals). 
When CDM isocurvature is introduced at $f_{\rm CDM} = 0.01$, the limit relaxes to $m_{\rm WDM} > 0.27~\mathrm{keV}$ (95\%~credible intervals), reflecting a shallow degeneracy in which blue-tilted isocurvature fluctuations partially compensate for WDM suppression. These results provide new constraints on thermal WDM in the presence of CDM isocurvature fluctuations and quantify the extent to which such fluctuations can mask the small-scale signatures of light relics. 
\end{abstract}

\maketitle

\vspace{1cm}

\begingroup
\hypersetup{linkcolor=black}
\renewcommand{\baselinestretch}{1.26}\normalsize
\tableofcontents
\renewcommand{\baselinestretch}{2}\normalsize
\endgroup

\newpage

\section{Introduction}\label{sec:intro}

The standard cosmological model, $\Lambda$CDM, has been shown  to successfully describe the universal large-scale structure (LSS). In $\Lambda$CDM, dark matter is assumed to be cold and dissipationless, interacting primarily through gravity. However, at smaller scales, several tensions have been revealed within the cold dark matter picture, often termed as small-scale structures (SSS) crisis (see, e.g., \cite{Bullock:2017xww, DelPopolo:2016emo, Weinberg:2013aya}, for a review). While the dwarf-galaxies problem has been partially alleviated by the discovery of newly identified faint and ultra-faint dwarf galaxies in Milky Way \cite{DELVE:2025jqe,Homma:2023ppu}, other problems persist. For instance, observations of faint and ultra-faint dwarf galaxies naturally raise the question why their subhalos fail to host enough stars to produce higher luminosity, called too big to fail problem \cite{Boylan-Kolchin:2011qkt,Boylan-Kolchin:2011lmk}. Moreover, observations of dwarf galaxies reveal that their innermost density distributions exhibit a cored density profile in contrast to the cuspy profile predicted by cold-dark-matter-only N-body simulations \cite{deBlok:2001,deBlok:2009sp}. 

Various proposals have been put forward to address the SSS issues within the CDM paradigm by incorporating tidal effects and baryonic processes such as active galactic nucleus and supernova feedback \cite{Wadepuhl:2010ib,Parry:2011iz,Governato:2012fa}.  Alternative dark-matter models have also been investigated extensively \cite{CosmoVerse:2025txj}. Among the possibilities, warm dark matter (WDM) provides an attractive mechanism to suppress small-scale structural growth \cite{Bode:2000gq,1982Nature}. WDM consists of semi-relativistic particles characterized by a free-streaming length $\lambda_{\rm FS}$, which describes the typical comoving distance a particle can travel freely before becoming gravitationally bound within potential wells. In this work, we restrict our discussion to WDM relics with an approximately \textit{thermal} Fermi-Dirac phase-space distribution. On scales larger than $\lambda_{\rm FS}$, WDM exhibit clustering properties akin to those of CDM. On smaller scale, free-streaming motion erases small-scale fluctuation, resulting in flattening of the otherwise cuspy inner density profiles and reduction in the abundance of the low-mass dark matter halos. 

Constraints on pure WDM have been derived across a wide array of astrophysical probes, including Lyman-$\alpha$ forest \cite{Irsic:2023equ}, milky way satellite counts \cite{Tan:2024cek, Nadler:2021dft,Dekker:2021scf}, strong gravitational lensing \cite{Gilman:2019nap}, luminosity function of high-redshift galaxies \cite{Menci:2016eui,Rudakovskyi:2021jyf}. A joint analysis using Lyman-$\alpha$, strong gravitational lensing and milky way satellite have disfavored $m_{\rm WDM}\lesssim 6 \;$keV at 2$\sigma$ confidence level \cite{Enzi:2020ieg}. However, previous studies such as in \cite{Maccio:2012qf} have shown that WDM particles with masses $\mathcal{O}(100) \;$eV are required to create significant cored halo profiles of radius $\mathcal{O}(\rm{kpc})$ for dwarf galaxies ($M\sim 10^{10} M_\odot$). The same work points out that WDM candidates of masses greater than $\rm{keV}$ fail to yield core sizes larger than about a kiloparsec in dwarf galaxies, thereby inhibiting their formations in the first place. These arguments and the current mass constraints have cast doubts on the viability of the WDM candidates as solutions to the SSS problems.

From a theoretical standpoint, nothing forbids thermal WDM masses well above the $\mathrm{keV}$ scale. Indeed, heavier thermal relics are perfectly consistent with structure formation and remain attractive dark-matter candidates: they retain the characteristic small-scale power suppression of WDM, but on scales smaller than those currently probed by Lyman-$\alpha$, strong lensing, or high-redshift galaxy counts. In this sense, WDM masses in the multi-keV range do not alleviate the small-scale structure tensions, but they remain fully viable and can be motivated by astrophysics and particle physics, especially in scenarios where WDM arises as a thermal relic or from minimal extensions of the Standard Model \cite{Covi:2009pq,Strumia:2010aa,FAYET1977461,Dodelson:1993je,Shi:1998km,Jaramillo:2022mos}.

While conventional analyses exclude thermal WDM masses below the keV scale in purely adiabatic scenarios, the possibility of viable sub-keV WDM remains scientifically interesting. As shown in a recent work \cite{Tadepalli:2025gzf}, even a small admixture of cold dark matter isocurvature (CDI) with a strongly blue-tilted power spectrum can partially compensate for the free-streaming suppression characteristic of light WDM, thereby reopening parameter space down to $m_{\rm WDM}\sim 600\,\mathrm{eV}$ without conflicting with CMB or BAO constraints. Access to this sub-keV regime by means of isocurvature is appealing for several reasons. First, WDM particles in the $\mathcal{O}(100)\,$eV mass range possess large free-streaming lengths capable of producing $\mathcal{O}(\mathrm{kpc})$ cores in dwarf-galaxy halos without invoking extreme baryonic feedback. Second, a sub-keV WDM particle imprints a qualitatively distinct scale of suppression, on $\sim 100\,\text{--}\,1000\,\mathrm{kpc}$ scales, creating a testable intermediate regime between galactic and Lyman-$\alpha$ constraints. Third, such light warm relics are natural outcomes of several non-thermal production mechanisms (e.g., sterile-neutrino-like species \cite{Shi:1998km,Dodelson:1993je} or axionlike warm relics \cite{Strumia:2010aa}), often accompanied by richer early-universe phenomenology. Lastly, thermal relics in the sub-keV mass range offer a significant cosmological advantage: matching the observed abundance requires only moderate effective degrees of freedom at decoupling ($g_{\rm *s} (T_{\rm dec})$), whereas keV-scale thermal WDM generally demands unrealistically large $g_{\rm *s} (T_{\rm dec})$ (\cite{Bode:2000gq,Paduroiu:2021cvg}. See Tab.~I in \cite{Vogel:2022odl}) or substantial nonstandard entropy dilution (\cite{Evans:2019jcs}). This makes sub-keV WDM considerably easier to realize in thermal scenarios. Together, these features make the prospect of viable sub-keV WDM particularly compelling, motivating careful exploration beyond the minimal adiabatic framework. Finally, a blue-tilted isocurvature mode by compensating WDM suppression, can decouple the characteristic suppression scale from the WDM mass, thereby allowing more freedom within the WDM-models. 

In this work, we examine the model proposed by~\cite{Tadepalli:2025gzf}, which introduces a mixed warm-dark-matter plus cold-dark-matter‐isocurvature (WCDM+CDI) framework. Specifically, we adopt their scenario in which a thermal WDM relic suppresses small-scale power via free-streaming, while an extremely sub-dominant CDM component carries blue-tilted isocurvature fluctuations that inject compensatory small-scale power. We embed this model into a full cosmological inference pipeline that combines large-scale data (CMB, BAO and SNe) with a small-scale dataset: the ultraviolet luminosity function (UVLF) of high-redshift galaxies, drawn from \textit{HST} and \textit{JWST} observations covering $4 \le z \le 11$. By leveraging the UVLF, we effectively extend our constraining reach to wavenumbers up to $k \sim \mathcal{O}(10)\,\mathrm{Mpc}^{-1}$, probing intermediate to small scales where the WCDM+CDI interplay is most active. With this joint dataset, we aim to answer the question of whether the data allow sub-keV WDM masses once the compensatory CDI freedom is introduced. In practice, we perform a Bayesian marginalization over the CDI amplitude and spectral index along with all other nuisance and cosmological parameters, thereby quantifying whether viable parameter space remains for $m_{\rm WDM}<1\,$keV under the extended WCDM+CDI hypothesis.

The remainder of this paper is organized as follows. In Sec.~\ref{sec:theory} we introduce the WCDM+CDI framework and outline its theoretical foundations. Sec.~\ref{sec:hmf} discusses the halo mass function (HMF) as a nonlinear probe of small-scale structure and presents illustrative examples of HMFs generated within the mixed WCDM+CDI scenario. In Sec.~\ref{sec:uvlf} we describe the implementation of the UVLF observable and its integration into our cosmological inference pipeline. Sec.~\ref{sec:mcmc_setup} details our Bayesian MCMC setup, including the likelihood components and baseline cosmological assumptions. Our main results are presented in Sec.~\ref{sec:results}, followed by a summary and discussion in Sec.~\ref{sec:discussion}.

\section{The WCDM+CDI Model}\label{sec:wcdm+CDI}

In this section, we describe the theoretical framework of the two-component warm and cold dark matter model with the cold dark matter species additonally carrying isocurvature fluctuations. This is followed by a brief discussion of the resulting halo mass functions. Our analysis focuses on fermionic WDM particles with a momentum distribution resembling that of a thermal relic. Their cosmological properties including free-streaming length, relic abundance, and velocity dispersion are fully specified by their mass and a Fermi-Dirac phase-space distribution.

\subsection{Theoretical setup}\label{sec:theory}

The theoretical details presented here largely follow the framework developed in the earlier work~\cite{Tadepalli:2025gzf}, and we provide below a concise outline relevant for the present analysis.

In the WCDM scenario, the total dark matter density is divided into a mixture of WDM and CDM components. The fractional contribution of the WDM component is defined as
\begin{equation}
    f_{\rm WDM} \equiv \frac{\Omega_{\rm WDM}}{\Omega_{\rm DM}} ,
\end{equation}
where $\Omega_{i}$ denotes the energy density parameter of component $i$. The CDM fraction is then given as $f_{\rm CDM} = 1-f_{\rm WDM}$.

The defining property of WDM is its free-streaming, which erases primordial fluctuations below the characteristic scale $\lambda_{\rm FS}$. On scales $k \gg k_{\rm FS}$, the matter power spectrum is therefore dominated by perturbations seeded by the CDM component, while the contribution from WDM is strongly suppressed. As structures begin to grow through gravitational collapse, the potential wells generated by CDM on small scales allow non-relativistic WDM particles to gradually fall in. However, the residual thermal velocities of WDM continue to free-stream, partially erasing these perturbations and thereby reducing the depth of the CDM potential wells. This leads to a coupled evolution: CDM tends to drag WDM into its gravitational wells, while the free-streaming of WDM suppresses the CDM transfer function relative to its early-time value. Consequently, the WDM transfer function is enhanced compared to a pure-WDM model, but its inefficient clustering still modifies the effective initial conditions for structure formation. The net effect is a smooth interpolation of the total matter transfer function, transitioning from the WDM cutoff scale to a suppressed CDM plateau at small scales. See Fig.~2 in \cite{Tadepalli:2025gzf}.

To discuss the matter power spectrum in the mixed WCDM models, it is often convenient to quantify deviations from the standard $\Lambda$CDM model by means of the \emph{transfer function}, defined as
\begin{equation}
    T^{\rm IC}_{X}(k) = 
    \sqrt{\frac{P^{\rm IC}_{X}(k)}{P^{\rm ad}_{\Lambda{\rm CDM}}(k)}} ,
\end{equation}
where $P^{\rm IC}_{X}(k)$ is the matter power spectrum of model $X$ with initial conditions IC denoting either adiabatic (ad) or a mixture of adiabatic and isocurvature perturbations (mx). Thus, $P^{\rm ad}_{\Lambda{\rm CDM}}(k)$ is the reference $\Lambda$CDM spectrum with purely adiabatic initial conditions.

To evaluate the matter power spectrum, one must solve the evolution of cosmological perturbations, governed by the coupled linearized Einstein and fluid equations, initialized with the primordial fluctuations. These are conventionally divided into two classes: \emph{adiabatic} fluctuations, which satisfy $\delta(n_j/s)=0$ for all species $j$ (with $n_j$ the number density and $s$ the entropy density), and \emph{isocurvature} fluctuations, characterized by $\delta(n_j/s)\neq 0$ at fixed total energy density. The primordial dimensionless power spectrum for either class is usually parameterized as a power law,
\begin{equation}
    \mathcal{P}_{i}(k) = A_{i}\left(\frac{k}{k_p}\right)^{n_{i}-1}, \label{eq:powerlaw}
\end{equation}
defined at a pivot scale $k_p=0.05\,{\rm Mpc}^{-1}$, where $A_{i}$ and $n_{i}$ are the amplitude and spectral index, respectively, for $i={\rm ad,\,iso}$.\footnote{We caution the reader that the spectral index $n_i$ used in Eq.~(\ref{eq:powerlaw}) should not be confused with the number density $n_j$. Henceforth, $n$ is used to denote the spectral index of the power spectrum.} Current CMB measurements~\cite{Planck:2018jri} indicate that the primordial fluctuations are predominantly adiabatic, with 
\[
A_{\rm ad}\simeq 2.1\times 10^{-9}, \qquad n_{\rm ad}\simeq 0.97.
\]
No significant detection of isocurvature has been made, although weak hints remain consistent with uncorrelated CDI models, with $A_{\rm cdi}\lesssim 0.6\, (0.04) A_{\rm ad}$ at $2\sigma$ confidence~\cite{Planck:2018jri} for scale-free (scale-invariant) shapes.

In the WCDM+CDI framework, thermal WDM produces a strong suppression of small-scale power, while a blue-tilted CDM isocurvature component can restore power on those scales. As a result, sub-keV WDM particles, otherwise excluded due to excessive suppression, may remain viable when compensated by CDI. This compensating effect can be quantified using the matter transfer function. The isocurvature degree of freedom is introduced in the CDM sub-component rather than in the warm relic itself for both physical and phenomenological reasons. From a dynamical standpoint, isocurvature fluctuations can easily act as a compensating source of small-scale structure if they are carried by a cold, pressureless species whose perturbations grow efficiently after horizon entry. By contrast, isocurvature modes carried by the WDM species are strongly damped by free-streaming. Although they are not erased entirely, and sufficiently steep primordial spectra (e.g.\ $n_{\mathrm{iso}} \gtrsim 4$--$5$) can in principle restore some small-scale power, the required tilts are considerably larger than those needed in the CDI case. Thus, a mixed WCDM+CDI scenario is a minimal extension in which isocurvature perturbations can meaningfully influence the small-scale matter power spectrum. A dedicated analysis of WDM-carried isocurvature, and the parameter space in which such compensation becomes viable, is left to future work.

In linear theory of LSS, the matter power spectrum for the WCDM+CDI model with mixed initial conditions can be expressed as
\begin{equation}
    P_{\rm WCDM}^{\rm mx}(k) = P_{\rm WCDM}^{\rm ad}(k) + P_{\rm WCDM}^{\rm cdi}(k)
\end{equation}where the superscripts mx(ad)(cdi) refer to the underlying mixed (adiabatic-only)(isocurvature-only) initial conditions.
The corresponding transfer function was derived in Ref.~\cite{Tadepalli:2025gzf} as
\begin{align}
    T_{\rm WCDM}^{\rm mx}(k) &\approx 
    \bigg[ 
        \left(f_{\rm WDM}T^{\rm ad}_{\rm WDM}(k) 
            \right.  \notag \\
        &\qquad \left.+ \left(1-f_{\rm WDM}\right)C(k)\right)^2 \notag \\
        &\qquad  
        + \left( \frac{r_c f_{\rm cdi}}{3} 
        R^{\rm cdi}_{\rm CDM}(k)C(k)\right)^2
    \bigg] ^{1/2}
\label{eq:Tk_WCDM_CDI}
\end{align}
where:
\begin{itemize}
  \item $f_{\rm WDM}$ is the warm dark matter fraction,
  \item $T^{\rm ad}_{\rm WDM}(k)$ is the adiabatic WDM transfer function,
  \item $r_{c}=\tfrac{(1-f_{\rm WDM})\Omega_{\rm DM}}{\Omega_{m}}$ is the CDM energy-density fraction,
  \item $f_{\rm cdi}=\sqrt{\frac{A_{\rm cdi}}{A_{\rm ad}}}$ is the CDM isocurvature fraction measured at a pivot scale $k_p$,
  \item $R^{\rm cdi}_{\rm CDM}(k)$ is the CDM isocurvature response function.
  \item $C(k)=\left[T^{\rm ad}_{\rm WDM}(k)\right]^{\beta f_{\rm WDM}}$ is the CDM correction factor introduced in Ref.~\cite{Tadepalli:2025gzf} which accounts for the suppression of the CDM potential well due to WDM free-streaming. Through empirical fitting, they found $\beta\approx 0.125$ as a suitable choice for the WCDM+CDI model with mixed initial conditions.
\end{itemize}

Eq.~\eqref{eq:Tk_WCDM_CDI} shows that the mixed WCDM+CDI transfer function depends on the isocurvature sector through the product $r_c f_{\rm cdi}$. Since $r_c$ is entirely fixed by the background cosmology, the impact of isocurvature perturbations on structure formation is controlled by the effective combination
\[
    r_c f_{\rm cdi} \propto (1 - f_{\rm WDM}) f_{\rm cdi}.
\]
For convenience, and to make explicit the degeneracy between the CDM fraction and the fractional isocurvature amplitude, we define the invariant isocurvature strength
\begin{equation}
    f_{\rm iso} \equiv (1 - f_{\rm WDM})\, f_{\rm cdi}.
    \label{eq:fiso}
\end{equation}
This quantity measures the actual contribution of CDM isocurvature to the total matter perturbation, independent of how the isocurvature amplitude $f_{\rm cdi}$ is assigned within the dark-matter sector. Larger values of $f_{\rm iso}$ correspond to stronger compensation of WDM free-streaming by CDM-carried isocurvature, whereas reducing the CDM fraction $(1 - f_{\rm WDM})$ suppresses the isocurvature impact even if $f_{\rm cdi}$ is held fixed. Defining $f_{\rm iso}$ therefore isolates the physically relevant isocurvature parameter and clarifies how the mixed model interpolates between the pure-adiabatic and strongly isocurvature-dominated regimes.

The CDI response function, $R^{\rm cdi}_{\rm CDM}$ encodes the spectral dependence of the blue-tilted CDI and the overall transfer function shape. The empirical expression for $R^{\rm cdi}_{\rm CDM}$, applicable over a wide range of scales was given in Ref.~\cite{Elena_Pierpaoli_1999} as
\begin{equation} R^{\rm cdi}_{\rm CDM} (k) = \frac{T_{\rm cdi}(q)}{T_{\rm ad}(q)} \left(\frac{k}{0.05\,\rm{Mpc^{-1}}}\right)^{(n_{\rm cdi} -n_{\rm ad})/2}\label{eq:Tcdi}
\end{equation}where
\be q = \frac{k}{\Omega_{\rm DM}h^2} {\rm Mpc}\ee and $T_{\rm cdi(ad)}(q)$ are obtained after smoothing over the BAO oscillations.\footnote{
For CDM isocurvature, the fitting functions are obtained as \cite{Elena_Pierpaoli_1999}
\begin{align*}
T_{\text{cdi}}(q) &= \left(1 + \frac{(40q)^2}{1 + 215q + (16q)^2 (1 + 0.5q)^{-1}} + (5.3q)^{8/5} \right)^{-5/4} \,, \\
T_{\text{ad}}(q) &= \frac{\ln(1 + 2.34q)}{2.34q} \\
& \qquad  \times \left(1 + 3.89q + (16.1q)^2 + (5.46q)^3 + (6.71q)^4 \right)^{-0.25} \,.
\end{align*}
}

The transfer functions for the pure WDM models with an adiabatic initial condition are well described by the following analytical expression (\cite{PhysRevD71,Bode:2000gq,Viel:2011bk,Abazajian:2005gj})
\begin{equation}
    T^{\rm ad}_{\rm WDM}(k,z=0) = \left[1 + (\alpha k)^{2\nu}\right]^{-5/\nu}
\end{equation}
where $\nu$ and $\alpha$ are derived from empirical fits to the numerical data, largely obtained from linear Boltzmann solvers. Following Ref.~\cite{Vogel:2022odl}, we write
\begin{align}
\label{eq:alpha_mWDM}
\alpha (m_{\rm WDM}) &= a \,
\left(\frac{m_{\rm WDM}}{1\ \mathrm{keV}}\right)^{b} \notag \\
& \qquad \times 
\left(\frac{\Omega_{\rm WDM}h^2}{0.12}\right)^{\eta}
\left(\frac{h}{0.6736}\right)^{\theta} 
\, h^{-1} \mathrm{Mpc},
\end{align}
where $a = 0.0437$, $b = - 1.188$, $\eta = 0.2463$, $\theta = 2.012$, and $\nu =1.049$ for spin-1/2 WDM particles, with a $10\%-$accuracy on scales $k\lesssim 2\alpha ~\rm{h/Mpc}$ for WDM masses $\gtrsim \mathcal{O}(0.5)$ keV.

\begin{figure*}[t]
    \centering
    \includegraphics[width=0.48\textwidth]{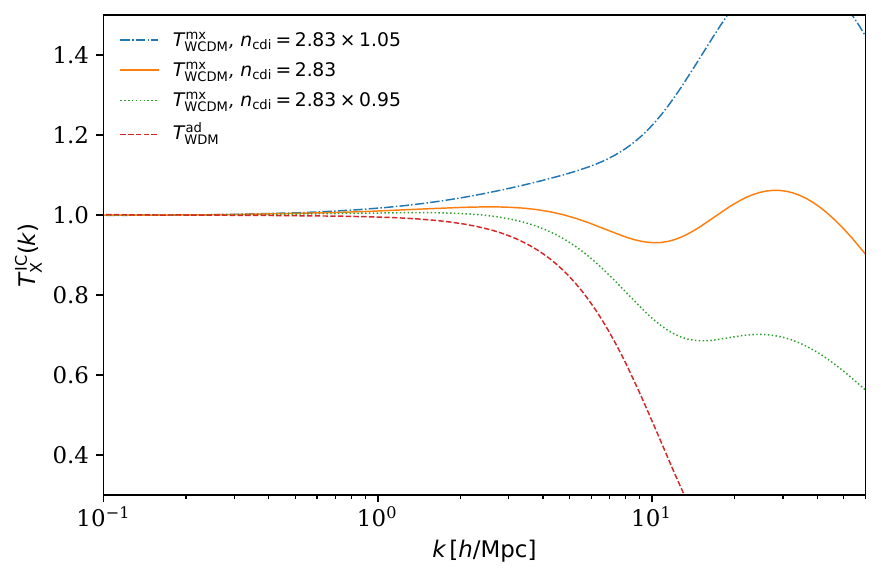}
    \includegraphics[width=0.48\textwidth]{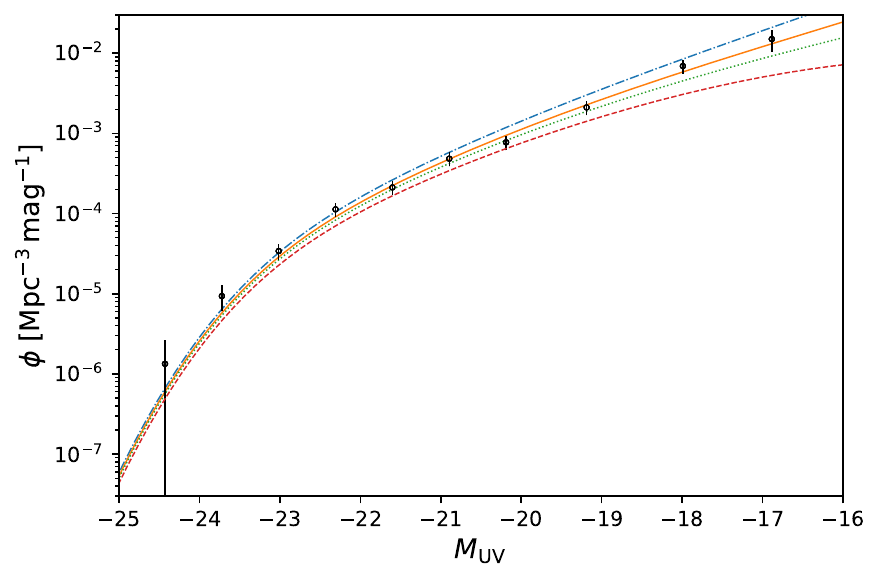}
    
\caption{\textbf{Left:} Linear matter transfer functions at $z=0$ from \textsf{CLASS} \cite{Blas:2011rf} for a WDM model with $m_{\rm WDM}=1~{\rm keV}$, $f_{\rm WDM}=0.99$ and three WCDM models with mixed adiabatic+CDI initial conditions ($f_{\rm iso}=0.293$, $n_{\rm cdi}=2.69,\,2.83,\,2.97$; green dotted, orange solid, blue dot-dashed). The red dashed curve shows the corresponding pure WDM adiabatic case, illustrating how a blue-tilted CDI component can partially offset WDM small-scale suppression.
\textbf{Right:} Representative UVLF curves at $z=6$ for the four cosmological models whose transfer functions are shown on the left.
}
    \label{fig:TWCDMiso}
 
\end{figure*}

In Fig.~\ref{fig:TWCDMiso}, we illustrate the compensating effect of CDI on the total matter power spectrum for a representative case with $m_{\rm WDM}=1\,{\rm keV}$ and $f_{\rm WDM}=0.99$, demonstrating how isocurvature power from even a tiny $1\%$ CDM fraction can counteract the suppression induced by WDM free streaming. In the same figure, we also present the corresponding theoretical UVLF curves at $z=6$ (with all nuisance parameters held fixed), overlaid on the \textit{HST} observational data. As discussed previously in \cite{Tadepalli:2025gzf} and highlighted here, introducing a blue-tilted CDM-isocurvature component can partially offset the free-streaming suppression of WDM power at intermediate scales. However, because the two transfer functions have intrinsically distinct scale dependencies, the cancellation is never exact, leaving behind residual oscillatory features. If these oscillations are sufficiently pronounced, they can be in conflict with small-scale data, leading to the rejection of regions of the WCDM+CDI parameter space despite the apparent average-level compensation.

Before turning to discuss halo mass functions, we note that several well-motivated early-universe mechanisms are known to produce isocurvature spectra with large blue-tilts ($n_{\rm cdi}\approx 3\,\text{--}\,4$):
\begin{itemize}
\item \textit{Axion isocurvature with Hubble-induced mass term:}
In \cite{Kasuya:2009up}, the authors construct a SUSY-embedded PQ spectator field model whose effective mass during inflation receives a Hubble-induced contribution from K\"{a}hler metric. When the PQ symmetry is spontaneously broken during inflation and the dominant radial field rolls from a large vev (owing to a flat-direction), the resulting spectral index of the axionic Goldstone mode is \(n_{\rm cdi} \approx 4 - 2\sqrt{9/4-(m/H)^2}\), where $m$ is the characteristic mass of the dominant radial field. For $m^2\gtrsim 2H^2$, the spectrum can behave as \(P_{\rm iso}(k)\propto k^{2\,\text{--}\,3}\) over the relevant small-scale regime.

\item \textit{Rotating complex scalars in the conformal limit:}  
A complex scalar field undergoing rotational motion during inflation can enter a 
conformal scaling regime when the radial mode rolls along a quartic potential with the scaling behavior $\propto a^{-1}$. This dynamics naturally produces a strongly blue isocurvature spectrum with $n_{\rm cdi}\!\simeq\!3$, corresponding to 
$P_{\rm iso}(k)\!\propto\!k^2$. This unified description of rotational and 
conformal-limit behavior was developed in \cite{Chung:2024ctx}.
\item \textit{Post-inflationary mechanisms:}
These include mechanisms such as spatial variations in primordial black hole abundance \cite{Afshordi:2003zb}, field fluctuations from symmetry breaking after inflation \cite{Enander:2017ogx}, and entropy modes
generated during phase transitions \cite{Elor:2023xbz}, 
and all yield spectra that rise steeply at low $k$ (with $n_{\rm cdi} \sim 4$) before rolling over at a characteristic scale set by causal physics.
\end{itemize}

\subsection{Halo mass function}\label{sec:hmf}
As elucidated above, the linear matter power spectrum provides the simplest window into the interplay between WDM free streaming and CDM isocurvature contributions. Ultimately it is the scales smaller than $\sim \mathcal{O}(0.1)$Mpc that yield observationally accessible signatures to constrain WDM. At these scales, the clustering of matter is typically non-linear for redshifts $z \lesssim 10$. The astrophysical quantity, halo mass function (HMF), serves as a crucial link between the primordial power spectrum and the abundance of collapsed (nonlinear) dark matter halos across cosmic time. Because the HMF is exponentially sensitive to changes in the underlying small-scale power, even modest differences in how WDM suppression and CDI compensation balance out can lead to significant shifts in the predicted halo abundances.

In particular, the residual oscillatory features left behind by the imperfect cancellation between WDM and CDI transfer functions can propagate into the HMF, producing scale-dependent modulations in halo counts. This makes the HMF a powerful diagnostic: by comparing predicted halo abundances with observations from galaxy surveys or strong lensing counts, one can discriminate between models that may appear degenerate at the level of the smoothed linear power spectrum. Thus, studying the HMF provides a natural next step toward constraining the viable parameter space of mixed WCDM+CDI scenarios.

The halo mass function quantifies the comoving number density of dark matter halos as a function of their mass. Formally, it is defined as
\[
\frac{dn}{d\ln M},
\]
which gives the number of halos per unit comoving volume per logarithmic mass interval. Since halos are the hosts of galaxies and larger cosmic structures, the HMF serves as the key statistical link between the underlying matter power spectrum and a wide range of astrophysical observables.

A well-known empirical expression for the HMF can be written as
\begin{equation}
    \tilde{n}(M)\equiv \frac{dn}{d\ln M} = -\frac{\bar{\rho}}{M}\, f(\nu)\, \frac{d \ln \sigma}{d\ln M},\label{eq:HMF}
\end{equation}
where:
\begin{itemize}
    \item $\bar{\rho}$ is the mean matter density of the universe,
    \item $\sigma(M)$ is the variance of the density field smoothed on mass scale $M$,
    \item $\nu = \delta_c/\sigma(M)$ is the peak-height parameter, and
    \item $f(\nu)$ is the multiplicity function, which encapsulates the collapse probability of density fluctuations.
\end{itemize}The halo mass associated with a smoothing scale
$R$ is given by
\begin{equation} 
M(R) = \frac{4\pi}{3}\bar{\rho}(c R)^3,
\end{equation}
 where $R$ denotes the Lagrangian radius enclosing mass $M$ at the mean density, for a convention-dependent mass-assignment factor $c$ \cite{Schneider:2013ria}.

The general structure of this expression arises from the Press-Schechter (excursion set) framework~\cite{Press:1973iz}, while the choice of $f(\nu)$ is calibrated either from analytic approximations (e.g.\ Press-Schechter, Sheth-Tormen~\cite{Sheth:2001dp} (ST)) or from numerical simulations (e.g.\ Tinker et al.~\cite{Tinker:2008ff}). In Eq.~(\ref{eq:HMF}), $\delta_c \simeq 1.686$ is the critical linear overdensity for spherical collapse.

Because $\sigma(M)$ is directly determined by the linear matter power spectrum, any modification to small-scale clustering, such as the suppression from WDM free streaming or the compensating isocurvature contribution from CDI, propagates into the HMF. This makes the HMF a particularly sensitive probe of models where linear power spectra may appear partially degenerate but still yield distinct predictions for halo abundances. 

In the excursion set ellipsoidal collapse framework~\cite{Sheth:2001dp}, the multiplicity function is given by
\begin{equation}
    f(\nu) = A \sqrt{\frac{2q\nu^2}{\pi}} \left[ 1 + (q\nu^2)^{-p} \right] e^{-q\nu^2/2}.
\end{equation}
The Sheth-Tormen parameters $p$, $q$, and $A \equiv A(p)$ are in general cosmology-dependent. Ref.~\cite{Gavas:2022iqb} demonstrated that the HMF of Eq.~(\ref{eq:HMF}) is not strictly universal, but instead exhibits an explicit dependence on the underlying power spectrum. Moreover, the best-fit parameters $p$ and $q$ show a mild dependence on the power-law spectral index of the primordial fluctuations. For the standard $\Lambda$CDM cosmology with a top-hat window function, the corresponding values are $p \simeq 0.3$, $A \simeq 0.3222$, and $q \simeq 0.707$ with $c=1$.

\begin{table*}[t]
\centering
\renewcommand{\arraystretch}{1.3}
\begin{tabular}{|p{3cm}|p{2cm}|p{2.5cm}|p{2.8cm}|p{3.0cm}|}
\hline
Model & $f_{\rm WDM}$ & $m_{\rm WDM}$ [eV] & initial conditions & $n_{\rm cdi},~f_{\rm iso}$ \\
\hline
$\Lambda$CDM        & 0.0 & --   & Adiabatic & -- \\
WDM-3000               & 1.0 & 3000  & Adiabatic & -- \\
m631          & 0.8 & 631  & Mixed     & 2.7,\;0.3464 \\
m251          & 0.9 & 251  & Mixed     & 3.1,\;0.6708 \\
m316          & 0.9 & 316  & Mixed     & 3.4,\;0.2236 \\
m501          & 0.7 & 501  & Mixed     & 2.9,\;0.4472 \\
\hline
\end{tabular}
\caption{Summary of cosmological models. $f_{\rm WDM}$ is the warm dark matter fraction, $m_{\rm WDM}$ is the WDM particle mass, and initial conditions denotes the type of initial conditions. For mixed initial-condition models, the spectral tilt $n_{\rm cdi}$ and effective isocurvature fraction $f_{\rm iso}$ are also given.}
\label{tab:model_summary}
\end{table*}

\begin{figure*}
    \centering
    \includegraphics[width=0.65\linewidth]{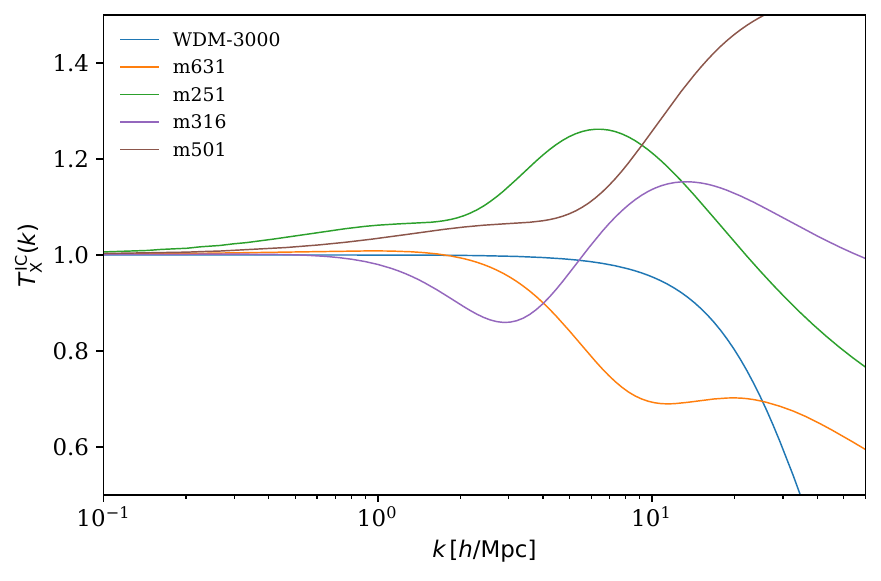}
    \caption{Plot of matter transfer function $T^{\rm IC}_{\rm X}(k)$ for models $\rm X$ with initial conditions $\rm IC$, as summarized in Tab.~\ref{tab:model_summary}. The solid blue curve shows a pure WDM model with particle mass $3 \,\mathrm{keV}$ and adiabatic initial conditions. The other curves correspond to WCDM+CDI models, where both WDM and CDM particles are present, with the CDM component carrying additional isocurvature fluctuations.}
    \label{fig:mPk_samples}
\end{figure*}

Below, we analyze the shape of the HMF predicted by a representative set of cosmological models. To construct this set, we randomly sampled several combinations of $m_{\rm WDM}$,  $f_{\rm WDM}$, $n_{\rm cdi}$ and $f_{\rm iso}$, from which four models are selected to illustrate the impact of these variations. To provide benchmarks, we also include a standard 
$\Lambda$CDM model and a pure WDM model with $m_{\rm WDM}=3~$keV. The six models considered are summarized in Tab.~\ref{tab:model_summary}, where only the listed parameters are varied while all other background cosmological parameters are held fixed. The benchmark $\Lambda$CDM and pure WDM models illustrate the generic HMF shapes typically constrained by UVLF datasets. Current WDM analyses using \textit{HST} or \textit{JWST} datasets suggest a $2\sigma$ lower bound
of $m_{\rm WDM}\gtrsim 2\,{\rm keV}$~\cite{Maio:2022lzg,Dayal:2023nwi}. A more recent estimate of $\sim3\,\rm{keV}$ in \cite{Liu:2024edl} obtained using the UVLF data is likely optimistic, as it includes simplified modeling assumptions such as utilization of a semi-empirical analytic fit for the HMF of thermal WDM relics and lack of any cosmic-variance in the dataset. In contrast, we propagate a conservative \(20\%\) cosmic-variance uncertainty floor following GALLUMI \cite{Sabti:2021xvh}, 
compute the linear matter power spectrum \(P(k)\) for each model with \textsf{CLASS}, and derive the HMF directly from this \(P(k)\) (rather than semi-empirical WDM fits), yielding \(m_{\rm WDM}\gtrsim 1.8\,\mathrm{keV}\) 
 at 2$\sigma$ when marginalized over all cosmological parameters. Beyond these benchmarks, we further examine four mixed WCDM+CDI models with different combinations of initial conditions and WDM particle masses, chosen such that they generate sizable residual oscillatory features of order $0.2-0.4$ in the matter transfer function. The corresponding transfer functions are shown in Fig.~\ref{fig:mPk_samples}.

\begin{figure*}[t]
    \centering
    \includegraphics[width=1\linewidth]{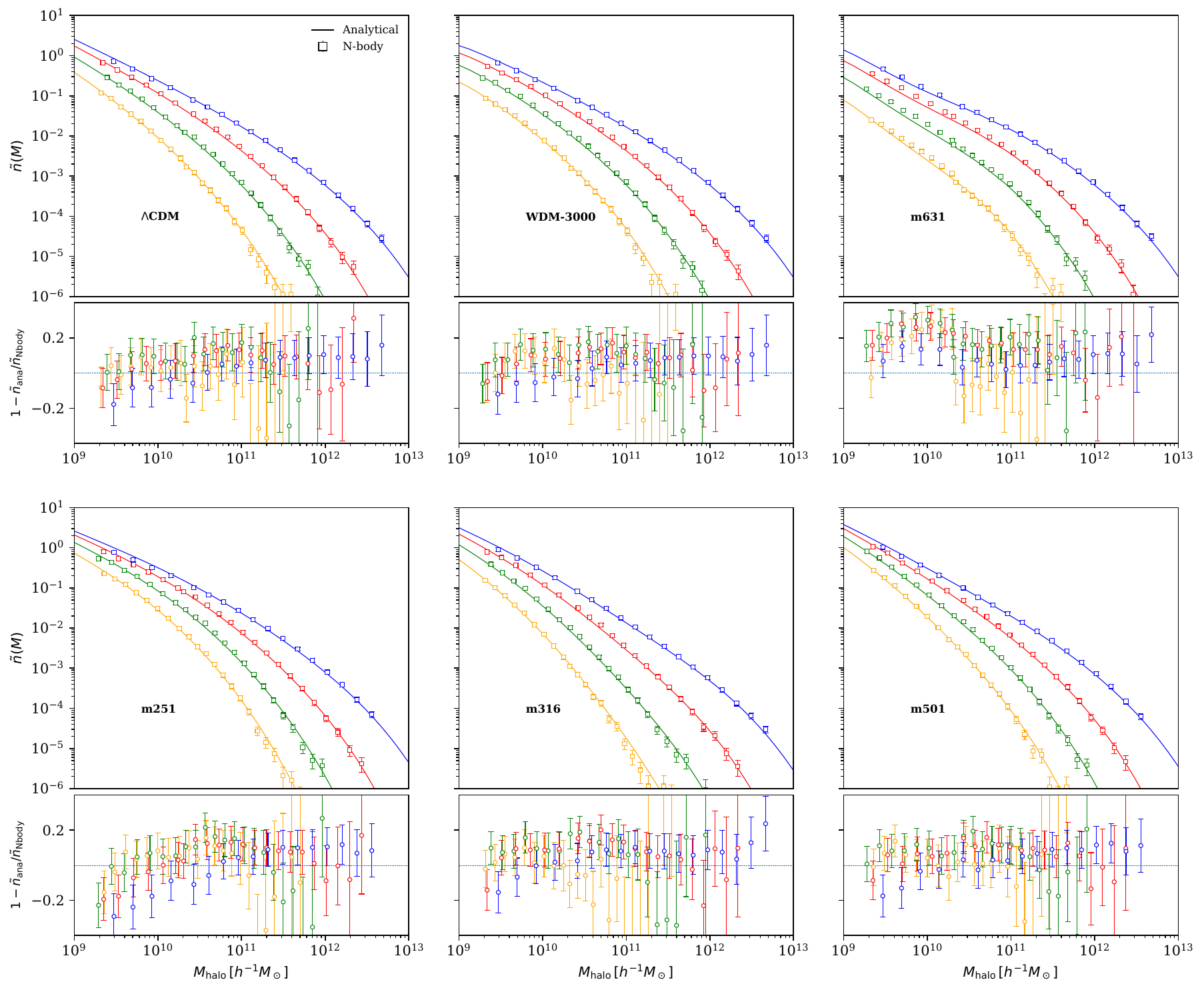}
    \caption{Halo mass functions (denoted as $\tilde{n}(M)$) for the six cosmological models at redshifts of  $z=4.0$ (in blue), $z=6$ (in red), $z=8$ (in green) and $z=10$ (in yellow), across mass-range from $10^{9}\,\text{--}\,10^{13}~M_\odot$.
    The square markers denote the $N$-body result using FoF algorithm, while the solid curves represent the theoretical predictions using the excursion-set theory. The WDM masses, $f_{\rm WDM}$ and the initial conditions for the various models are listed in Tab.~\ref{tab:model_summary}. The analytical ST expression reproduces the simulation results to within $\sim 20\%$ accuracy across nearly all models and redshifts considered.
    }
    \label{fig:hmf_nbody}
\end{figure*}

In Fig.~\ref{fig:hmf_nbody}, we compare the HMFs obtained from DM-only $N$-body simulations with the analytical prediction from Eq.~\eqref{eq:HMF} adopting the ST form.\footnote{We perform DM-only N-body simulations using \textsf{FastPM} \cite{Feng:2016yqz} which is an approximated
particle mesh N-body solver. The output phase space data from \textsf{FastPM} is processed using \textsf{nbodykit} \cite{Hand:2017pqn} to generate the halo mass function.} Results are shown for the six representative cosmological models summarized in Tab.~\ref{tab:model_summary}, with HMFs evaluated at redshifts $z=4,~6,~8,$ and $10$, across the halo mass range $10^{9}$–$10^{13}~M_\odot$, which encompasses the UVLF scales relevant to our analysis. The square markers indicate simulation measurements based on the Friends-of-Friends (FoF) algorithm~\cite{osti_5014740}. In addition to the intrinsic Poisson uncertainty on the simulated results, we include cosmic variance as discussed above. The solid curves represent the analytical ST predictions. We find that a single sharp-$k$ filter tuned to reproduce the small-scale behavior of WDM models systematically overpredicts the abundance of massive halos in regimes where the linear matter power spectrum follows the standard CDM-like behavior. Consequently, for the \textsf{WDM-3000} and \textsf{m631} models, which exhibit a pronounced small-scale suppression, we compute the HMF using a hybrid  filtering approach: a sharp-$k$ filter with parameters $A=0.35442$, $p=0.3$, $q=0.86$, and $c=2.4$ at low masses, smoothly transitioned to a tophat filter at high masses. This prescription yields good agreement with the simulations across the full redshift and mass ranges without requiring the additional ellipsoidal corrections suggested by \cite{Schneider:2013ria}. The final results presented in this work are not significantly sensitive to the details of this filtering procedure. 

Overall, the analytical ST expression reproduces the simulation results to within $\sim 20\%$ accuracy across nearly all models and redshifts considered. We have further verified that this level of agreement persists when using alternative fitting choices, such as the ST form with a smooth filter or alternative parameterizations such as Tinker et al. 
 \cite{Tinker:2008ff}. 
The residual scatter reflects the intrinsic theoretical uncertainty of analytical HMF prescriptions, which must be taken into account in addition to the standard cosmic-variance uncertainty when comparing with UVLF measurements. To ensure that these theoretical and cosmic-variance contributions 
are consistently incorporated, we impose a conservative $20\%$ uncertainty floor on all UVLF data points as explained in the next section.

\section{UV Luminosity Function}\label{sec:uvlf}

The galaxy ultraviolet luminosity function measures the comoving number density of galaxies per unit luminosity, $\Phi_{\rm UV}=dn_g/dM_{\rm UV}$.\footnote{The luminosity in the UV band is primarily dominated by the hot, short-lived O- and B-bright stars within a galaxy.} It provides one of the most direct observational probes of galaxy formation in the high-redshift universe, and, through its connection to the halo mass function, a sensitive lever on cosmological parameters. In this work we employ the GALLUMI likelihood from Ref.~\cite{Sabti:2021xvh} for analyzing the UVLF data. Since the original GALLUMI is restricted to only \textit{HST} dataset, we adapt it to account for additional \textit{JWST} dataset. We also make subtle changes to ST HMF modeling in order to provide a good fit for the WCDM+CDI models. In what follows we describe the UVLF data sets used, the construction of the likelihood, and the role of astrophysical nuisance parameters and systematic error treatments.

\subsection{UVLF data}
We compile UVLF measurements of galaxies from \textit{HST} at $z\simeq 4\,\text{--}\,8$ \cite{Oesch_2018,Bouwens_2021} and from \textit{JWST} at $z\simeq 9\,\text{--}\,11$~\cite{donnan2024jwst}, spanning magnitudes from the bright exponential cutoff ($M_{\rm UV}\sim -22$) to the faint-end power-law tail ($M_{\rm UV}\sim -16$). Higher redshift UVLF datasets from \textit{JWST} are limited by noise and Poisson statistics and do not improve the bound on WDM substantially. Each datum consists of a central redshift, an absolute magnitude bin (with effective bin width $\Delta M$), an estimate of the number density $\Phi$ in units of ${\rm Mpc}^{-3}{\rm mag}^{-1}$, and associated uncertainties. For \textit{HST} at $z=4,5$ and $6$, the statistical errors are typically $\sim 10\%$, whereas \textit{JWST} measurements at redshifts greater than $10$ are more noise-dominated, often exceeding $30\%$.

An important subtlety concerns whether reported observational errors already include field-to-field fluctuations (cosmic variance). In many UVLF catalogs,  the published error bars reflect only Poisson (counting) statistics, in which case an additional error component must be added to represent cosmic variance. To account for the cosmic variance and the theoretical uncertainty in analytic ST expression for the HMF, we adopt a conservative $20\%$ floor on the error of each data point, corresponding to the expected magnitude of cosmic variance in deep surveys with areas of order $\sim 100~{\rm arcmin}^2$. Following the prescripton in \cite{Sabti:2021xvh}, we conservatively adopt the larger of the two contributions:
\begin{equation}
\sigma_i^2 = \max\left(\sigma_{{\rm data},i}^2 , \big(0.2\,\Phi_{{\rm data},i}\big)^2 \right).\label{eq:CV}
\end{equation}
This conservative prescription ensures that field-to-field variance is not neglected, particularly for deep, narrow surveys where it dominates over Poisson noise.

\subsection{From cosmology to the UVLF}
Our forward model proceeds from cosmological initial conditions to observed galaxy luminosity function in a sequence of steps, adapted from \cite{Sabti:2021xvh}:
\begin{enumerate}
\item First, we evaluate the redshift-dependent linear matter power spectrum $P(k,z)$ for a given cosmological model using the linear Boltzmann solver such as \textsf{CLASS}.
\item We compute the variance of the matter density field on a given halo mass scale, $\sigma(M_h,z)$, from the matter power spectrum $P(k,z)$ using the hybrid filtering procedure described in Sec.~\ref{sec:hmf}. In this approach, $\sigma(M_h,z)$ is evaluated with a sharp-$k$ filter for halo masses $M_h < M_{\rm h, cut}$ and with a real-space tophat filter for $M_h > M_{\rm h, cut}$ 
where \footnote{Here \(M_{\rm h,cut}\) is chosen to scale with the WDM particle mass as \(M_{\rm h,cut} \propto m_{\rm WDM}^{-3.33}\), matching the standard scaling of the WDM half-mode mass with \(m_{\rm WDM}\) that follows from the dependence of the free-streaming wavenumber on the thermal relic mass. The normalization at \(m_{\rm WDM} = 3~\mathrm{keV}\) is a calibration choice.}
\be
M_{\rm h, cut}=10^{11}M_\odot \left(\frac{3\,\rm{keV}}{m_{\rm WDM}}\right)^{3.33} .
\ee
For the sharp-$k$ filter, we adopt the parameter values $A = 0.35442$, $p = 0.3$, $q = 0.86$, and $c = 2.4$, obtained from fits to the simulated halo mass functions across the range of models and redshifts shown in Fig.~\ref{fig:hmf_nbody}.

\item The halo mass function (HMF) is subsequently evaluated in the ``universal'' form 
\be
\frac{dn}{dM_h}
 = M_h^{-1} \frac{dn}{d \ln M_h}
\ee
where $\frac{dn}{d \ln M_h}$ is defined in Eq.~(\ref{eq:HMF}). Unless stated otherwise, we adopt the Sheth-Tormen parameterization. Using the $\sigma(M_h,z)$ obtained in the previous step from both the sharp-$k$ and tophat filters, we smoothly splice the corresponding HMFs at the transition mass scale $M_{\rm h, cut}$.

\item The halo–galaxy connection is modeled through the star-formation efficiency (SFE) $f_\star(M_h,z)$, defined such that the star-formation rate is given by $\dot{M}_\star(z) = f_\star(M_h,z)\,\dot{M}_h(z)$, modulo a double–power-law dependence on halo mass:
\begin{equation}
\label{eq:SFE}
    f_\star
    = \frac{\epsilon_\star(z)}{\left(M_h/M_c(z)\right)^{\alpha_\star(z)}
    + \left(M_h/M_c(z)\right)^{\beta_\star(z)}} ,
\end{equation}
where $\alpha_\star < 0$, $\beta_\star > 0$, $M_c > 0$, and $\epsilon_\star > 0$ are UVLF astrophysical nuisance parameters, each potentially redshift dependent. The instantaneous UV luminosity $L_{\rm UV}$ is related to the star-formation rate through a fixed conversion factor \cite{Kennicutt:1998zb,Madau:1996hu,Madau:1997pg}, $\dot{M}_\star(z) = \kappa_{\rm UV} L_{\rm UV}$ with 
\begin{equation}
\kappa_{\rm UV} = 1.15 \times 10^{-28}\; M_\odot\,\mathrm{s}\,\mathrm{erg}^{-1}\,\mathrm{yr}^{-1} \,.
\end{equation}
 Luminosity and absolute magnitude are connected via
\[
\log_{10}\!\left(\frac{L_{\rm UV}}{\mathrm{erg}\,\mathrm{s}^{-1}}\right)
= 0.4\,\left(51.63 - M_{\rm UV}\right).
\]
See \cite{Sabti:2021xvh} for additional details.

\item Our baseline results adopt model~II from GALLUMI, in which the halo accretion rate is prescribed as
\begin{equation}
    \dot{M}_h(z)
    = c\, M_h\, H(z),
\end{equation}
where the proportionality constant $c$ is absorbed into the SFE normalization parameter $\epsilon_\star$. Differences among commonly used halo-accretion prescriptions primarily modify the halo-mass and redshift scaling through simple power-law factors. Any dependence of $\dot{M}_h(z)$ in the form of $M_h^p$ is degenerate with the parameters $\alpha_\star$ and $\beta_\star$ in the star-formation rate $\dot{M}_\star(z)$, while redshift-dependent contributions of the form $H(z)^q$ are absorbed by the redshift evolution of the internal nuisance parameters.

\item The mapping $M_h \mapsto M_{\rm UV}$ from above yields the intrinsic, scatter-free UVLF (assuming one central galaxy in
each halo)
\begin{equation}
\Phi(M_{\rm UV},z) = \frac{dn}{dM_h}\,\frac{dM_h}{dM_{\rm UV}} .
\end{equation}
\item The intrinsic astrophysical scatter/stochasticity in the mass-luminosity relation is modeled as a Gaussian of width $\sigma_{M_{\rm UV}}$, convolved with the above expression, which broadens the distribution and especially boosts number densities at the bright end (Eddington bias)~\cite{eddington_formula_1913}.
\item Finally, the UVLF likelihood follows a Gaussian $\chi^2$ form:
\begin{equation}
-2\ln \mathcal{L} = \sum_{i,j}\left(\frac{\Phi_{\rm model}(z_i,M_{\mathrm{UV},j}) - \Phi_{\rm data}(z_i,M_{\mathrm{UV},j})}{\sigma_{i,j}}\right)^2 ,
\end{equation}
where the sum runs over all bins in redshift and magnitude. The error $\sigma_{i,j}$ includes the quoted observational uncertainty and a 20\% floor as shown in Eq.~(\ref{eq:CV}). 

Dust corrections are applied to the data using an empirical IRX-$\beta$ calibration as described in \cite{Sabti:2021xvh}. The effect is a horizontal shift in magnitudes and a rescaling of bin widths, introducing an additional source of systematic uncertainty in the overall normalization. Additionally, when comparing theory with data, we account for the Alcock–Paczyński (AP) effect. Since observed UVLFs are reported assuming a fiducial cosmology to convert angles and redshifts into comoving volumes and absolute magnitudes, we rescale the effective distances and survey volumes using the sampled cosmology. This correction leads to small but non-negligible shifts in number densities and ensures that cosmology enters consistently through both structure formation and observational projection.
\end{enumerate}

We note that the prefactor $A_{\rm ST}$ in the Sheth-Tormen (HMF formulation)  rescales the entire halo mass function. It represents a pure vertical scaling of the UVLF, independent of magnitude and redshift. As we discuss below, this parameter is partially degenerate with $\epsilon_\star$ and $\sigma_{\rm M_{\rm UV}}$.

The astrophysical nuisance parameter $\epsilon_\star$ shifts all halo luminosities by a constant factor, which in magnitude space corresponds to a horizontal translation
\begin{equation}
M_{\rm UV} \to M_{\rm UV} - 2.5\log_{10}\epsilon_\star.
\end{equation}
This is degenerate with a vertical rescaling only in the faint-end power-law regime, where the slope $d\ln \Phi/dM_{\rm UV}$ is constant. Near $M_c$ and bright end, where the slope varies rapidly, $\epsilon_\star$ produces shape changes that cannot be mimicked by $A_{\rm ST}$. Meanwhile, increasing scatter $\sigma_{M_{\rm UV}}$ broadens the halo-luminosity mapping, boosting bright-end counts (due to steep slopes) while having minimal effect at the faint end. This changes the shape rather than providing a uniform scaling.

Thus, while $A_{\rm ST}$, $\epsilon_\star$, and $\sigma_{M_{\rm UV}}$ all impact the overall normalization, they do so in distinguishable ways: $A_{\rm ST}$ vertically, $\epsilon_\star$ horizontally (with vertical imprint via the slope), and $\sigma_{M_{\rm UV}}$ through bright-end boosting. In practice, partial degeneracies exist (e.g.\ between $A_{\rm ST}$ and $\epsilon_\star$ if only faint-end data are used), but they can be broken with broad magnitude coverage. In our UVLF analysis where the \textit{HST} and \textit{JWST} data are limited to magnitudes within the range $-22\lesssim M_{\rm UV}\lesssim -16$, we find that $A_{\rm ST}$ is highly degenerate with the nuisance parameters $\epsilon_\star$ and $\sigma_{M_{\rm UV}}$, and thus any residual normalization mismatch between the analytic HMF and $N$-body/observed halo counts is absorbed within the UVLF parameters.

\section{MCMC setup}\label{sec:mcmc_setup}
In Sec.~\ref{sec:wcdm+CDI}, we noted that thermal WDM leads to a strong suppression of small-scale power (below $\lambda_{\rm FS}$) through free–streaming, whereas uncorrelated CDM isocurvature perturbations, if sufficiently blue–tilted, can inject additional power on exactly those scales. In Sec.~\ref{sec:uvlf}, we introduced the UVLF dataset along with its evaluation pipeline, which enables us to place cosmological constraints from small-scales within range $0.5~{\rm Mpc}^{-1} \lesssim k \lesssim 10~{\rm Mpc}^{-1}$. To constrain this mixed WCDM+CDI cosmology, we exploit the complementary reach of small-scale UVLF data and large-scale CMB+BAO+SNe measurements.

\subsection{Datasets and likelihoods}
The UVLF dataset as detailed in Sec.~\ref{sec:uvlf} constrains power over $0.5~{\rm Mpc}^{-1} \lesssim k \lesssim 10\,{\rm Mpc}^{-1}$, but is insensitive to larger scales 
$k \lesssim \mathcal{O}(0.1)~{\rm Mpc}^{-1}$.
CMB anisotropies, in contrast, provide some of the most stringent limits on large-scale cosmological parameters. Uncorrelated CDM isocurvature modes leave characteristic signatures in the CMB: they shift the acoustic phase, alter TT peak ratios, produce distinctive TE/EE features, modify the low-$\ell$ TT plateau via the early ISW effect, and even flip the sign of the TE correlation for entropy modes. These effects enable the CMB to strongly anchor both the CDI fraction and the standard cosmological parameters.

Our CMB likelihood follows the Planck 2018 release. We include \textsf{SRoll2} low-$\ell$ TT/EE likelihoods ($\ell \lesssim 30$) \cite{Pagano:2019tci}, the high-$\ell$ nuisance-marginalized \textsf{Plik} TT/TE/EE likelihood up to $\ell_{\max}=2500$ with a free calibration $A_{\rm Planck}$~\cite{Planck:2018lkk}, and the Planck PR3 lensing potential likelihood over $8 \leq L \leq 400$ \cite{Planck:2018lbu}. To further constrain geometry and late-time clustering, we include DESI Y1 BAO measurements \cite{DESI:2024uvr,DESI:2024lzq,DESI:2024mwx} and distance moduli measurements from SNe type-Ia supernovae from \textsf{Pantheon} \cite{Pan-STARRS1:2017jku,Jones:2017udy}.

In our analysis, we fit the UVLF using 69 data points drawn from \textit{HST} and \textit{JWST} measurements. Three magnitude bins, two at $z\simeq 9$ and one at $z\simeq 10$, all centered at $M_{\mathrm{UV}}\simeq -21$, were excluded from the main inference. These bins appear to behave as statistical outliers relative to the full UVLF sample and lie in a regime where observational systematics and completeness corrections are known to be largest. When they are included, the additional freedom in the isocurvature parameters $(n_{\mathrm{cdi}},\,f_{\mathrm{iso}})$, combined with flexibility in the UVLF nuisance parameters, enables the model to contort itself within a very small region of parameter space in an attempt to accommodate these outliers. This generates a sharply peaked, highly non-Gaussian posterior and drives several UVLF parameters to unphysical values—for example, pushing $\alpha_\star$ to $-1.2$ instead of the typical $\Lambda$CDM/$\Lambda$WDM value near $-0.6$, along with multi-$\sigma$ excursions in $M_c$ and $\epsilon_\star$. Simultaneously, the CDI parameters move into regions such as $n_{\rm cdi}>4.5$ which are outside of the usual cosmological priors. Such behaviour is characteristic of overfitting driven by a few high-leverage points rather than evidence for a genuine cosmological signal.

To avoid contaminating the cosmological inference with these outlier-driven distortions, we excluded the three problematic data-points and base our constraints on the remaining 69 points. This trimmed dataset yields a stable, physically reasonable posterior, recovers UVLF parameters consistent with $\Lambda$CDM/$\Lambda$WDM expectations, and shows no indication of a CDI component once the anomalous bins are removed.

\subsection{Sampler and pipeline}
We perform Markov-chain Monte-Carlo (MCMC) sampling using \textsf{Cobaya} \cite{Torrado:2020dgo}, fitting jointly to CMB, BAO, SNe and UVLF likelihoods. Since the original UVLF likelihood, GALLUMI, was developed for \textsf{MontePython}~\cite{brinckmann_montepython_2019}, we adapted and validated its implementation within \textsf{Cobaya}. Each run consists of four independent chains initialized from a pre-computed covariance matrix. Final constraints are derived from the converged chains using \textsf{GetDist}~\cite{lewis_getdist_2025}.

\subsection{Cosmology and prior}
Our fiducial cosmology is a spatially flat WCDM model with massless neutrinos and a cosmological constant $\Lambda$. In this mixed DM-framework, the warm fraction is fixed at $0.99$, thereby isolating the impact of a small $1\%$ CDM admixture. Our results remain robust for $f_{\rm CDM}$ values in the range $1\,\text{--}\,5\%$. Note that CDM fractions larger than $\sim10\%$ would naturally relax the lower bound on the WDM mass. All species are assumed to carry nearly scale-invariant adiabatic perturbations. In the WCDM+CDI model, we assign uncorrelated isocurvature initial conditions exclusively to CDM.

CMB anisotropy and matter power spectra are generated using \textsf{CLASS}. For CMB lensing, we restrict the calculation to the linear matter power spectrum to avoid inaccuracies reported in \cite{Chung:2023syw} for nonlinear fitting functions for large blue-tilts ($n_{\rm cdi}\gtrsim3.5$). This choice ensures that our resulting constraints are slightly conservative.

Within \textsf{CLASS}, primordial CDI is characterized by its amplitude \(f_{\rm cdi}\) at the pivot scale \(k_p\) and by its spectral index \(n_{\rm cdi}\), specified through a power-law form (without cutoff) as given in Eq.~(\ref{eq:powerlaw}). From the perspective of models that generate blue-tilted isocurvature fluctuations, however, one generally expects a cutoff scale beyond which the spectrum ceases to grow, either flattening or decreasing due to the underlying physics \cite{Chung:2015tha}. For simplicity, we assume that any such cutoff lies outside the range of small scales probed by the UVLF, so that a pure power-law description remains adequate for the present analysis.  

As discussed in Sec.~\ref{sec:wcdm+CDI}, CDI power spectra with spectral tilts \( n_{\rm cdi} \lesssim 4\) are readily motivated within cosmology, particularly through spectator-field dynamics during inflation \cite{Kasuya:2009up,Chung:2015tha,Chung:2024ctx} or through post-inflationary mechanisms such as those discussed in \cite{Afshordi:2003zb,Enander:2017ogx}.
By contrast, achieving tilts \( n_{\rm cdi} >4\) typically requires nonstandard dynamics, for instance brief resonant states or phase transitions can generate sharply peaked or bumpy features in the power spectrum rather than a sustained power law \cite{Chung:2021lfg,Chung:2023xcv,Chung:1999ve,Barnaby:2009dd}. Thus, while blue spectra with \(n_{\rm iso} \sim \mathcal{O}(1\,\text{--}\,4)\) arise naturally in multi-field or curvaton-like settings, substantially larger effective tilts generally indicate localized or exotic dynamics, with the enhanced power confined to at most $\mathcal{O}(1)$ decade in scale rather than extending over many decades as in a pure power law. Motivated by these considerations, we adopt a prior range  \(n_{\rm cdi}\in [0,4.5]\), thereby restricting our analysis to physically reasonable blue-tilted CDI models. The physically relevant measure of the matter isocurvature contribution is \(f_{\rm iso}\), defined in Eq.~(\ref{eq:fiso}). 

For the UVLF nuisance parameters, we assume them to be redshift independent. This simplifying choice reduces the astrophysical freedom in the galaxy–halo connection and therefore yields the most conservative (i.e., tightest) lower limits on the WDM particle mass. Allowing these parameters to vary with redshift would introduce additional degeneracies that can partially mimic the suppression of small-scale structure induced by WDM, thereby weakening the resulting constraints. By fixing the redshift evolution of the UVLF parameters, we ensure that any observed deficit (or excess) of faint galaxies is attributed primarily to the underlying cosmological parameters rather than to additional astrophysical degrees of freedom.

The full set of sampled and derived parameters, together with their associated priors, is summarized in Tab.~\ref{tab:mcmc_parameters}. The priors for the UVLF astrophysical parameters are motivated from $\Lambda$CDM runs.

\begin{table}[t] 
\centering \renewcommand{\arraystretch}{1.3} \begin{tabular}{|p{2.5cm}|p{2.5cm}|p{3cm}|} \hline Model & Parameter & Prior/Derived \\ 
\hline 
\textbf{WCDM} & $\Omega_b h^2$ & $\mathcal{U}[0.019,0.026]$ \\ 
 &$\Omega_{\rm DM}h^2$ & $\mathcal{U}[0.08,0.15]$ \\ 
 &$\tau_{\rm reio}$ & $\mathcal{U}[0.02,0.10]$ \\ 
 &$\theta_{\rm s,100}$ & $\mathcal{U}[1.035,1.045]$ \\  &$\log(10^{10} A_s)$ & $\mathcal{U}[2.6,3.4]$ \\ 
 &$n_s $ & $\mathcal{U}[0.92,1.05]$ \\ 
 &$1\rm{keV}/m_{\rm WDM}$ & $\mathcal{U}[0.01,1]$ \\ 
 &$\Omega_{\rm CDM}h^2$ & $\Omega_{\rm DM}h^2 (1-f_{\rm WDM})$ \\ 
 &$\Omega_{\rm WDM}h^2$ & $\Omega_{\rm DM}h^2 f_{\rm WDM}$ \\ 
\hline 
 \textbf{WCDM+CDI} &$1\rm{keV}/m_{\rm WDM}$ & $\mathcal{U}[0.1,8]$ \\ 
 &$n_{\rm cdi}$ & $\mathcal{U}[0.,4.5]$ \\ 
 &$f_{\rm iso}$ & $\mathcal{U}[0.,1]$ \\ 
 &$f_{\rm cdi}$ & $f_{\rm iso}/(1-f_{\rm WDM})$ \\ 
\hline
 \textbf{UVLF} &$\alpha_\star$ & $\mathcal{U}[-0.75,-0.35]$ \\ 
\textbf{parameters} &$\beta_\star$ & $\mathcal{U}[0.,3]$ \\ 
 &$\log_{10}\epsilon_{\star}$ & $\mathcal{U}[-3,0.5]$ \\ 
 &$\log_{10}M_{c}$ & $\mathcal{U}[10,14]$ \\ 
 &$\sigma_{\rm M_{\rm UV}}$ & $\mathcal{U}[0.001,3]$ \\ 
 \hline
\end{tabular}
\caption{Sampled and derived cosmological parameters for the baseline MCMC analysis. Uniform priors are denoted by $\mathcal{U}$. The WDM fraction is fixed at $f_{\rm WDM}=0.99$.}
\label{tab:mcmc_parameters}
\end{table}

\section{Results}\label{sec:results}
We now present the results for the WCDM+CDI model, constrained jointly by the CMB+BAO+SNe+UVLF likelihoods. For comparison, we include the WCDM model. This provides a direct benchmark against the standard thermal WDM mass limits, fully exploiting the small-scale constraining power of the UVLF measurements. Together, this suite of models allows us to disentangle the respective roles of small- versus large-scale datasets, to quantify the impact of CDI fluctuations.

\begin{figure*}[t]
    \centering
    \includegraphics[width=0.6\textwidth]{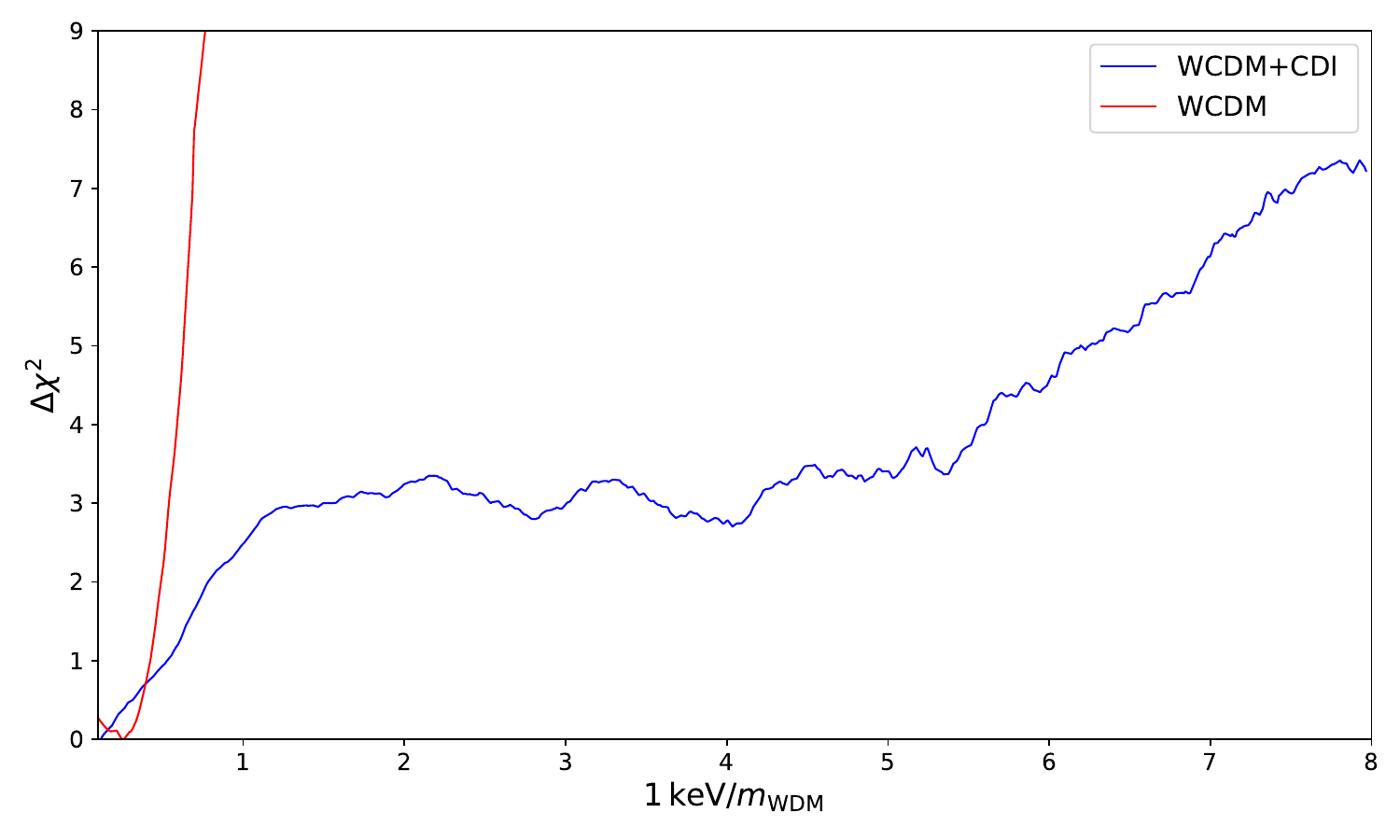}
    \caption{Profile likelihood for the WCDM and WCDM+CDI models, constrained using the CMB+BAO+SNe+UVLF likelihoods. The inclusion of CDM isocurvature fluctuations noticeably extends the allowed low-mass region of the WDM parameter space, flattening the profile likelihood down to masses as low as $ 200\,\mathrm{eV}$ and thereby weakening the lower bound on the WDM mass.}
    \label{fig:profile_lik}
\end{figure*}

\begin{figure*}
    {\centering
    \includegraphics[width=0.7\linewidth]{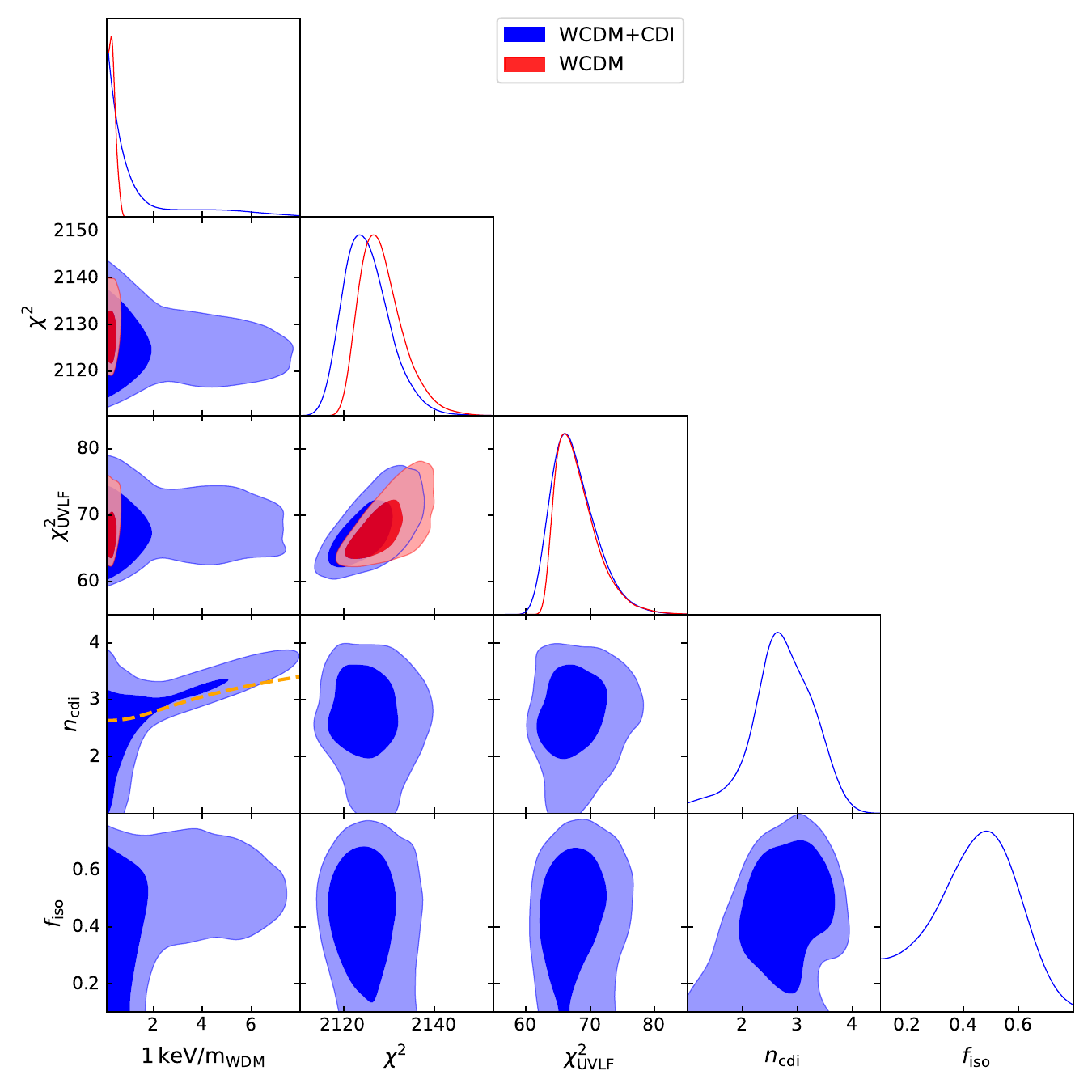}\par}
    \caption{One- and two-dimensional marginalized posterior distributions for the baseline WCDM and WCDM+CDI models, constrained using the CMB+BAO+SNe+UVLF likelihoods. The inclusion of CDM isocurvature fluctuations primarily broadens the 2$\sigma$ contour toward lower WDM masses, effectively compensating for the small-scale suppression induced by WDM and thereby extending the allowed parameter space. We also include  an approximate principal degeneracy direction between $n_{\rm cdi}$ and $m_{\rm WDM}^{-1}$ as a dashed orange curve using Eq.~(\ref{eq:ncdi_final}).
}
    \label{fig:2D_plot}
\end{figure*}

Our key results from the full CMB+BAO+SNe+UVLF likelihood analysis are as follows:
\begin{enumerate}
    \item For the baseline WCDM model (with no isocurvature), the 1- and 2-$\sigma$ lower limits on the WDM mass are $m_{\rm wdm} > 2.9~\mathrm{keV}$ and $m_{\rm wdm} > 1.8~\mathrm{keV}$, respectively.
    \item When allowing for CDM isocurvature fluctuations in the WCDM+CDI model, the corresponding 1- and 2-$\sigma$ bounds weaken to $m_{\rm wdm} > 1.7~\mathrm{keV}$ and $m_{\rm wdm} > 0.27~\mathrm{keV}$, respectively.
\end{enumerate} In both of these models, the marginalized constrain on the six cosmological parameters ($\omega_b,\,\omega_{\rm DM},\,\theta_{\rm s,100},\,A_s,\,n_s,\tau_{\rm reio}$) remain consistent with $\Lambda$CDM. The WDM fraction is fixed at $f_{\rm WDM}=0.99$.

In Fig.~\ref{fig:profile_lik}, the profile likelihood for the WCDM and WCDM+CDI models reveals that the inclusion of CDM isocurvature fluctuations significantly weakens the frequentist lower bound on the WDM mass. In the pure WCDM model, the likelihood rises steeply for $m_{\rm wdm}^{-1} \gtrsim 1\,{\rm keV}^{-1}$, excluding low WDM masses at high confidence. However, when CDI degrees of freedom are introduced, the profile likelihood develops a broad and nearly flat plateau, with $\Delta\chi^{2}\approx 3$ persisting down to effective masses as low as $\sim 200\,{\rm eV}$. This behavior reflects the fact that blue-tilted CDI fluctuations can compensate for the small-scale suppression induced by light WDM, permitting parameter combinations that would otherwise be ruled out by the UVLF data. The CDI--WDM compensation opens a shallow degeneracy direction in which the likelihood remains acceptable. Thus, from a frequentist standpoint, the resulting $95\%$ confidence lower limit becomes substantially weaker when marginalizing over CDI.

The Bayesian marginalized posteriors in Fig.~\ref{fig:2D_plot} reveal a similar qualitative behavior as the profile likelihood. Although the inclusion of CDI weakens the 1$\sigma$ lower bound on the WDM mass by roughly a factor of two, the 2$\sigma$ posterior shows much larger support for the extremely low WDM masses, and conseqeuntly the lower limit weakens approximately a factor of 7, almost consistent with the factor of $\mathcal{O}(10)$ predicted in \cite{Tadepalli:2025gzf}. Achieving successful compensation of the WDM-induced small-scale suppression at $m_{\rm wdm}\!\lesssim\!0.4$~keV requires a narrow region of parameter space in $(f_{\rm iso}, n_{\rm cdi})$: most combinations of CDI amplitude and tilt either overproduce or underproduce small-scale power and are strongly disfavored by the UVLF data. This comparison highlights a key complementarity. The likelihood surface reveals the existence of a compensation mechanism between WDM and CDI, whereas the Bayesian inference quantifies how strongly this mechanism is limited by the directional parametric freedom required for it to operate.

Within the region of parameter space corresponding to lighter sub-keV WDM masses, the posterior in Fig.~\ref{fig:2D_plot} tends to cluster around $n_{\rm cdi}\!\sim\!3\,\text{--}\,4$. In our convention 
$P_{\rm iso}(k)\propto k^{\,n_{\rm cdi}-1}$, this corresponds to a strongly 
blue-tilted isocurvature spectrum scaling approximately as $k^{2\,\text{--}\,3}$. While this behavior 
should not be interpreted as evidence for a physical isocurvature signal, part of the clustering arises from prior-volume effects and the particular parameterization of the CDI  sector, it is nevertheless fully consistent with theoretical expectations whenever the model attempts to allow sub-keV WDM masses. The principal CDI--WDM degeneracy direction is derived in Appendix~\ref{App:degeneracy} (see Eq.~\eqref{eq:ncdi_final}) using the analytical transfer function in Eq.~(\ref{eq:Tk_WCDM_CDI}). For clarity, we reproduce the resulting approximation here:
\begin{align}
    n_{\rm cdi} 
     \approx~ & 0.223 \ln \left(1.\, +0.177\left(\frac{1\,\mathrm{keV}}{m_{\rm WDM}}\right)^{2.49}\right) \notag \\
    & -0.377 \ln (f_{\rm iso})+ 2.33.
\end{align}
This is illustrated in Fig.~\ref{fig:2D_plot}, where we overlay the approximate principal degeneracy direction between $n_{\rm cdi}$ and $m_{\rm WDM}^{-1}$ as a dashed orange curve, evaluated at a representative mean value of $f_{\rm iso}\approx 0.45$.

In Fig.~\ref{fig:2D_plot}, we also note the tight distributions in $\chi^2_{\rm UVLF}$ which indicate that both models fit the UVLF data equally well. The data do not require additional flexibility beyond the baseline WDM model, but they do allow such freedom in the extended WCDM+CDI model by relaxing the constraints on smaller WDM masses. However, the overall $\chi^2$ shows a modest improvement, with $\Delta\chi^2 \approx -7$ for the WCDM+CDI model, primarily arising from a better fit to the CMB data enabled by the isocurvature freedom.

\begin{figure*}
    \centering
    \includegraphics[width=0.48\linewidth]{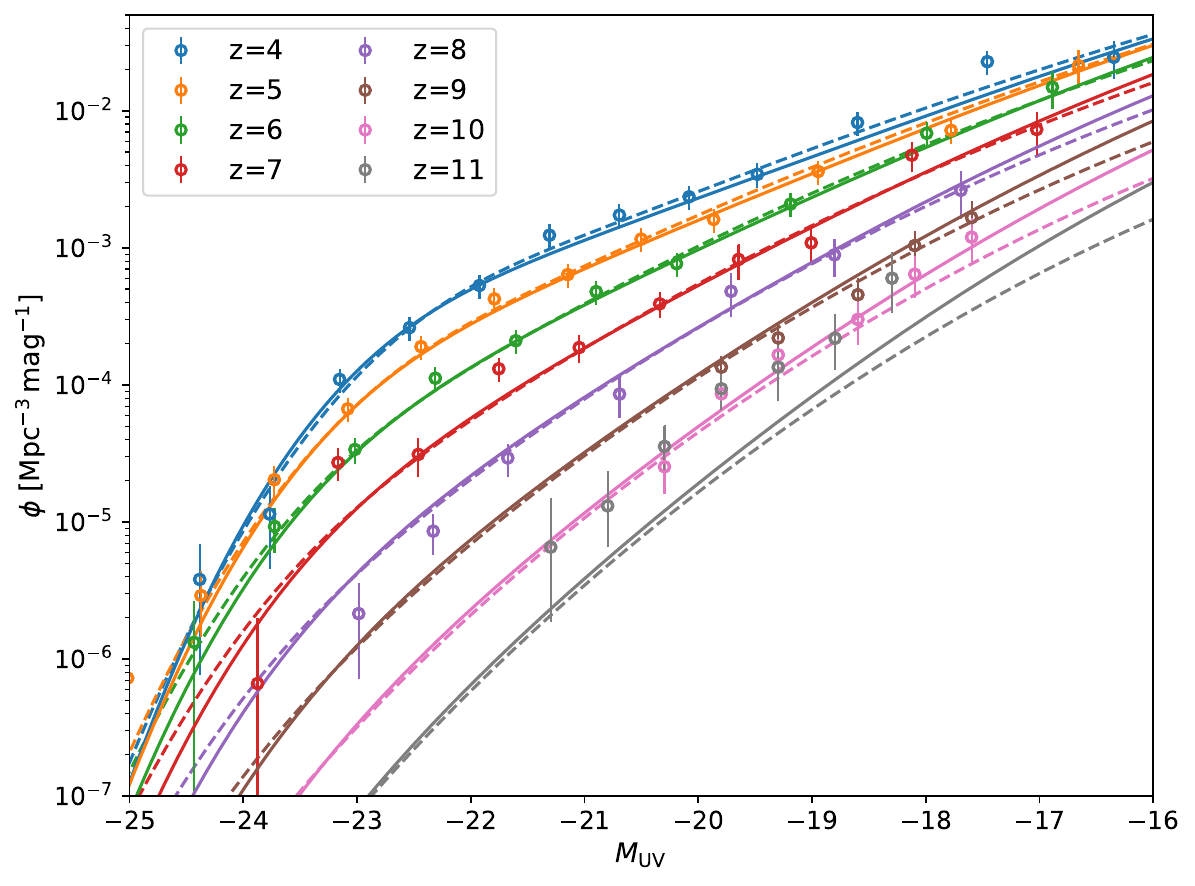}
    \includegraphics[width=0.48\linewidth]{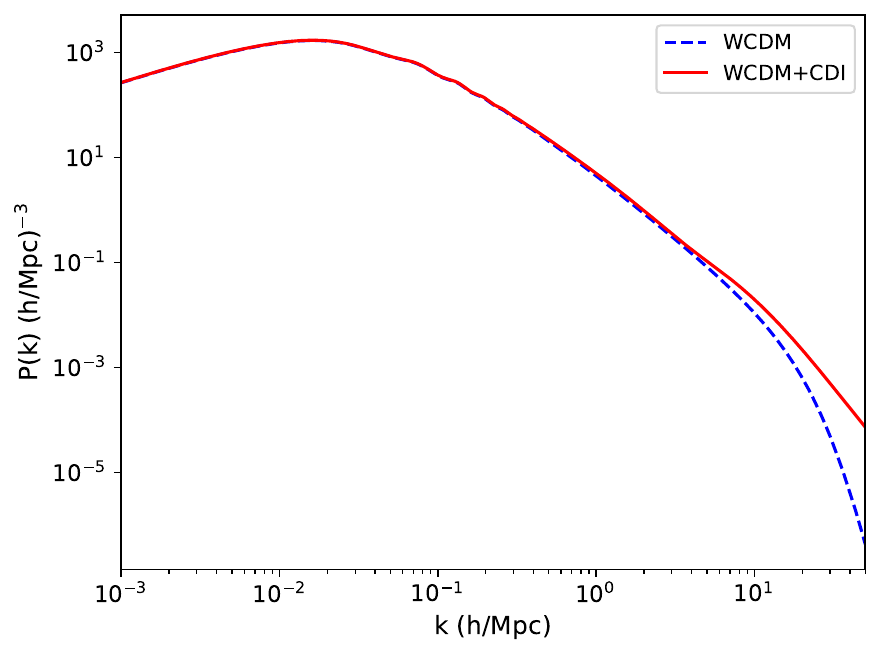}
    
\caption{
Comparison of UVLFs in the redshift range $4\,\text{--}\,11$  and matter power spectrum at $z=4$ between the WCDM and WCDM+CDI models using representative parameter sets at the 2$\sigma$ lower WDM-mass boundary. The solid curves show the WCDM+CDI prediction for $m_{\rm WDM}=405\,$eV with $n_{\rm cdi}=2.93$ and $f_{\rm iso}=0.61$, while the dashed curves show the baseline WCDM prediction for $m_{\rm WDM}=1920\,$eV. The figure illustrates how CDM isocurvature fluctuations allow the WCDM+CDI model to match the observed faint-end UVLF even at significantly lower WDM masses. The UVLF data points shown in this plot have been corrected for dust and AP effect.}
    \label{fig:uvlf_overlay}

\end{figure*}

To illustrate how the two models differ in their predictions for the observed UV luminosity function, we extract representative parameter sets located at the 2$\sigma$ lower-mass boundary of each model. For both WCDM and WCDM+CDI, we select the MCMC sample with the minimum $\chi^{2}$ value within a narrow window around the corresponding 2$\sigma$ $m_{\rm WDM}$ limit. Using these parameter sets, we compute the model UVLFs and compare them directly to the observational data. In Fig.~\ref{fig:uvlf_overlay}, the solid curve shows the prediction from the WCDM+CDI model for $m_{\rm WDM}=405\,$eV with $n_{\rm cdi}=2.93$ and $f_{\rm iso}=0.61$, while the dashed curve corresponds to the baseline WCDM for $m_{\rm WDM}=1920\,$eV with no CDM isocurvature. This comparison highlights how CDM isocurvature fluctuations enable the WCDM+CDI model to reproduce the observed faint-end counts even at substantially lower WDM masses.

Finally, from the best-fit parameters for a sub-keV WDM, the total matter isocurvature amplitude at smaller scales can be inferred from the
parameters \(n_{\rm cdi}\) and \(f_{\rm iso}\) as
\begin{equation}
    A_{\rm cdi}(k) \;\approx\;
    A_{\rm ad}(k_p)
    \left(\frac{k}{k_p}\right)^{n_{\rm cdi}-1}
    f_{\rm iso}^2 .
\end{equation}
For \(f_{\rm iso}\!\sim\!0.5\) and \(n_{\rm cdi}\!\sim\!3\,\text{--}\,4\), the resulting
amplitude at \(k\!\sim\!\mathcal{O}(100)\,h/\mathrm{Mpc}\) lies in the range
\(\mathcal{O}(10^{-3})~\text{--}~\mathcal{O}(1)\), well below current spectral–distortion
limits from COBE/FIRAS~\cite{Fixsen:1996nj} and may become marginally detectable in
future experiments such as PIXIE \cite{Kogut:2011xw}.

\section{Discussion}\label{sec:discussion}

We have tested whether a small CDM admixture carrying blue-tilted, uncorrelated isocurvature fluctuations can compensate the small-scale suppression induced by a sub-keV thermal warm dark matter (WDM) relic. Using a joint CMB+BAO+SNe+UVLF (HST + JWST) analysis and marginalizing over the CDI amplitude and tilt, our main findings are:
\begin{itemize}
\item In the pure adiabatic WDM scenario, our joint analysis yields a lower bound of \(\approx 1.8\,\mathrm{keV}\) on the thermal WDM particle mass at the \(95\%\) confidence level.
\item When CDM isocurvature fluctuations are included (WCDM+CDI), a shallow degeneracy emerges: a blue-tilted isocurvature spectrum \((n_{\rm cdi}\sim3\,\text{--}\,4)\) can partially compensate for WDM free-streaming, allowing significantly lighter thermal relics. In this extended model the \(95\%\) C.L. lower limit relaxes to $\approx 270$ eV.
\end{itemize}

Our conclusions rest on several simplifying choices that affect the quantitative bounds. The Bayesian constraints exhibit only mild sensitivity to the choice of priors, and the profile-likelihood and posterior distributions are consistent with each other. The UVLF modeling (star-formation efficiency, scatter, halo-galaxy mapping) and the adopted \(20\%\) cosmic-variance floor are important systematic inputs. Relaxing the redshift-independence of nuisance parameters would weaken constraints. The HMF calibration used a hybrid analytic filter tuned to DM-only simulations. 

The remaining compensated parameter space can be constrained more decisively by combining orthogonal small-scale probes that are sensitive to the residual spectral features left by imperfect CDI--WDM cancellation. In particular, Lyman-$\alpha$ forest flux-power measurements reach higher wavenumbers and involve astrophysical systematics distinct from those of the UVLF, making them especially powerful for testing small-scale structure. Strong-lensing substructure, through flux-ratio anomalies and measurements of the subhalo mass function, directly probes the abundance of low-mass halos and can therefore reveal or exclude the spectral modulations characteristic of compensated models. Complementary constraints arise from Milky Way satellites and stellar streams: satellite counts and stream-gap analyses are sensitive to local subhalo populations and to any departures from the standard CDM-like abundance at masses below $\sim 10^{9}\,M_\odot$. In addition, 21-cm cosmology during cosmic dawn and reionization provides a highly sensitive probe of the timing and abundance of the earliest small halos, offering an independent test of suppressed or compensated small-scale power. Finally, deeper and wider UVLF measurements from forthcoming \textit{JWST} and Roman surveys will probe a substantially larger comoving volume and multiple independent sightlines, thereby reducing sample (cosmic) variance and sharpening constraints on the faint-end population. This improved control of both cosmic variance and Poisson noise will further restrict the viable region of WCDM+CDI parameter space.

We recommend several directions for future work. First, since the surviving compensated parameter space can in principle yield sufficiently cored halo profiles, dedicated DM-only and hydrodynamic simulations initialized with mixed WCDM+CDI initial conditions would provide essential study of the halo mass function, subhalo statistics, and internal halo structure. Second, as emphasized above, a joint multi-probe program combining UVLF measurements with Lyman-$\alpha$ forest data, strong-lensing substructure, Milky Way satellite counts, stellar streams, and forthcoming 21-cm observations would offer a decisive test of the compensated scenario across complementary scales and redshifts. In parallel, our results motivate theoretical work aimed at constructing explicit and more \emph{natural} inflationary or spectator-field models capable of generating the required blue isocurvature spectrum while remaining consistent with CMB large-scale isocurvature and non-Gaussianity constraints. Finally, we note that the compensated-framework approach explored here is not restricted to thermal WDM: the same methodology can be applied to other scenarios that suppress small-scale power, including nonthermal warm relics, interacting or dissipative dark-matter models, and other beyond-$\Lambda$CDM mechanisms with characteristic cutoffs.

In conclusion, a blue-tilted isocurvature initial condition, in addition to the standard nearly scale-invariant adiabatic mode, provides a physically motivated and testable mechanism for partially masking the small-scale suppression characteristic of thermal WDM. The residual spectral features arising from imperfect cancellation render the scenario empirically falsifiable or favorable with more observations and dedicated simulations. While CDI introduces additional model flexibility, it nonetheless reopens a sizable region of viable parameter space for light, sub-keV thermal WDM.

\section*{Acknowledgement}
We thank Nashwan Sabti for helpful discussions and clarifications regarding the implementation and interpretation of the GALLUMI likelihood. This work was supported by the U.S. Department of Energy under Grant No.~DE-SC0025611 (RC). This research was supported in part by Lilly Endowment, Inc., through its support for the
Indiana University Pervasive Technology Institute, and by JSPS KAKENHI Grant Numbers 25K01004 (TT) and MEXT KAKENHI 23H04515 (TT), 25H01543 (TT).

\begin{appendix}

\section{Derivation of the CDI--WDM degeneracy direction}
\label{App:degeneracy}

In this Appendix, we derive the analytic degeneracy direction between a blue-tilted CDI component and WDM free-streaming that is used in the main text. The starting point is the mixed-transfer function for the WCDM+CDI model from Eq.~(\ref{eq:Tk_WCDM_CDI}),
\begin{align}
  T_{\rm WCDM}^{\rm mx}(k) &\approx 
   \bigg[ 
   \left(f_{\rm WDM}T^{\rm ad}_{\rm WDM}(k) 
      \right.  \notag \\
     &\qquad \left.+ \left(1-f_{\rm WDM}\right)C(k)\right)^2 \notag \\
       &\qquad  
     + \left( \frac{r_c f_{\rm cdi}}{3} 
         R^{\rm cdi}_{\rm CDM}(k)C(k)\right)^2
        \bigg] ^{1/2}
    \label{eq:Tmx_repeat}
    \end{align}

For sub-keV WDM, the adiabatic WDM transfer function is strongly suppressed on small scales and, for the small-scale regime probed by the UVLF, the leading adiabatic term in \eqref{eq:Tmx_repeat} becomes negligible compared with any CDI-sourced contribution. In the limit where the WDM adiabatic piece vanishes, the mixed transfer function is dominated by the CDI term and
\[
T_{\rm mx}(k) \simeq \frac{r_c\,f_{\rm cdi}}{3}\;R^{\rm cdi}_{\rm CDM}(k)\;C(k).
\]

Our degeneracy direction is defined by the condition that the transfer function $T_{\rm mx}(k)$ over an extended range of small-sclaes proved by UVLF shows no sign of suppression comapred to the $\Lambda$CDM.  Concretely, we impose
\begin{equation}
\frac{r_c\,f_{\rm cdi}}{3}\;R^{\rm cdi}_{\rm CDM}(k)\;C(k) \Big|_{k = k_\star} \approx 1,
\label{eq:deg_cond}
\end{equation}
with the scale choice,
\[
k_\star \simeq 10 \rm{/Mpc},
\]
which represents a typical smallest comoving wavenumber to which the UVLF is sensitive. Setting the left-hand side of \eqref{eq:deg_cond} to unity is equivalent to solving
\begin{equation}
\ln\!\Biggl[\Bigl(\tfrac{r_c\,f_{\rm cdi}}{3}\,R^{\rm cdi}_{\rm CDM}(k_\star)\,C(k_\star)\Bigr)^2\Biggr] = 0,
\label{eq:log_condition}
\end{equation}
which produces a linear equation for $n_{\rm cdi}$ once all other factors are evaluated at $k_\star$.

Next, we evaluate the logarithm in \eqref{eq:log_condition} by substituting the power-law dependence of the CDI response function. We consider the following fiducial parameter values:
\[
h = 0.6736,\quad \Omega_{\rm DM} = 0.266,\quad \Omega_m = 0.315,
\quad n_{\rm ad}\simeq 0.97,
\]
and the WDM-fit parameters $a,b,\eta,\theta,\nu$ given in the main text below Eq.~(\ref{eq:alpha_mWDM}). We also adopt the fiducial fractions used in the numerical exploration, $f_{\rm WDM}=0.99$ and $f_{\rm CDM}=0.01$, so that $r_c = f_{\rm CDM}(\Omega_{\rm DM}/\Omega_m)$. Evaluating the transfer-function ratio $\bigl[T_{\rm cdi}(q_\star)/T_{\rm ad}(q_\star)\bigr]$ and the correction factor $C(k_\star)$ at $k_\star=10 /{\rm Mpc}$ produces purely numerical coefficients. Carrying through the algebra (numerical evaluation of the $k$-dependent factors followed by solving the linear equation for $n_{\rm cdi}$) yields the compact relation
\begin{align}
n_{\rm cdi} \simeq \, & 
 0.223\,
\ln\!\left(1 + 0.177\left(\frac{\rm{1\,keV}}{m_{\rm WDM}}\right)^{2.49}\right) \notag \\
& -0.377\,\ln(f_{\rm iso}) + 2.33,
\label{eq:ncdi_final}
\end{align}
which is shown as the orange line in Fig.~\ref{fig:2D_plot}.

\subsection*{Interpretation and limitations}

Equation~\eqref{eq:ncdi_final} provides an approximate degeneracy direction in the $(m_{\rm WDM},n_{\rm cdi},f_{\rm iso})$ parameter space along which a CDI component can compensate the small-scale suppression produced by WDM free-streaming at the characteristic scale $k_\star$. A number of caveats apply:

\begin{enumerate}
  \item The additive constant $2.33$ and the numerical coefficients in Eq.~\eqref{eq:ncdi_final} depend on the choice of evaluation scale $k_\star$, the fiducial cosmological parameters, and the particular empirical fitting formulas used for $T_{\rm cdi}(q)$, $T_{\rm ad}(q)$ and $T^{\rm ad}_{\rm WDM}(k;m)$. Changing $k_\star$ or the cosmology will modify these numbers, though the qualitative dependence on $\ln(1+0.177/m^{2.49})$ and $\ln(f_{\rm iso})$ is robust.
  \item The degeneracy condition we impose is that the CDI piece alone reaches unit amplitude in Eq.~\eqref{eq:deg_cond}. This is an operational choice motivated by the fact that for sub-keV WDM the adiabatic-WDM contribution is negligible on the small scales probed by the UVLF; equivalently one may impose other conditions suitable for all mass range.
  \item All transfer-function expressions employed are empirical fits to Boltzmann-solver outputs and are valid within their published ranges (see main text for references and validity ranges). The formula \eqref{eq:ncdi_final} should therefore be used as a guide to the degeneracy direction rather than a substitute for a full forward-modeling pipeline when performing precision parameter inference.
\end{enumerate}

The above relation provides a practical one-parameter description of how a blue CDI spectrum (tilt $n_{\rm cdi}$) must be adjusted as a function of WDM mass $m_{\rm WDM}$ and isocurvature amplitude $f_{\rm iso}$ in order to reproduce the same small-scale transfer amplitude as $\Lambda$CDM at the UVLF-probed scale. The same derivation may be repeated for alternative cosmological probes at different scales.

\begin{figure*}[t]
    \centering
    \includegraphics[width=0.8\linewidth]{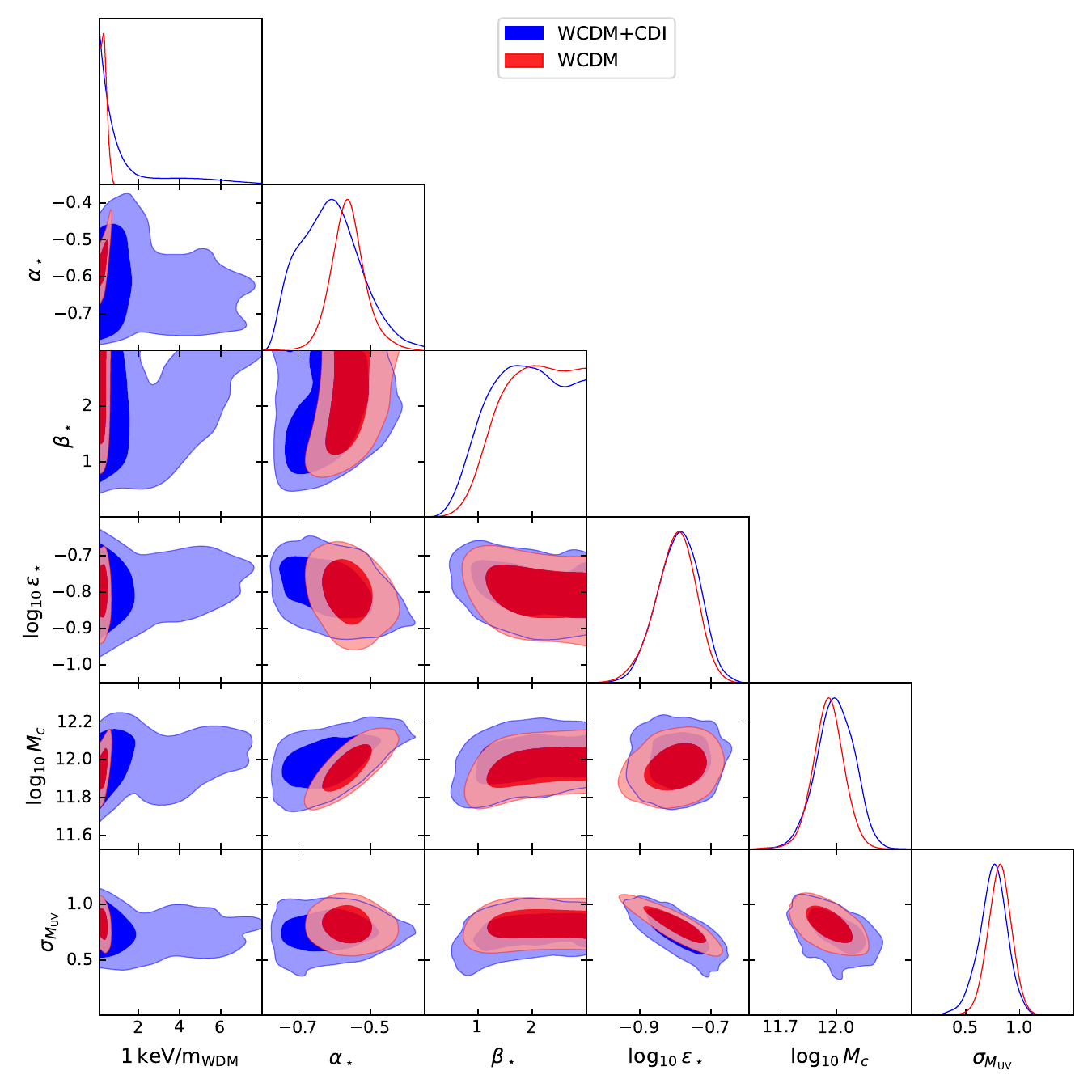}
\caption{
Posterior distributions for the UVLF nuisance parameters.
}
    \label{fig:uvlf_nuisance}
\end{figure*}

\section{Posterior plot for the UVLF parameters}

In Fig.~\ref{fig:uvlf_nuisance}, we show the posterior distributions for the UVLF nuisance parameters for the WCDM and WCDM+CDI models. In the WCDM+CDI model, the preferred values of $\alpha_\star$ shift slightly lower for sub-keV WDM masses than in the baseline WCDM case because the blue-tilted CDM isocurvature component supplies additional small-scale power, reducing the need for steep faint-end star-formation efficiency. For the remaining UVLF parameters, the posterior profiles from the two models are comparatively similar. 

\end{appendix}

\pagebreak
\nocite{apsrev41Control}
\bibliographystyle{JHEP2.bst}
\bibliography{newref,refs}

\end{document}